\documentclass{article}

\usepackage{fullpage}
\usepackage{cite}
\usepackage{amsmath,amssymb,amsfonts, mathtools}
\usepackage{graphicx}
\usepackage{textcomp}
\usepackage[dvipsnames]{xcolor}
\usepackage{tikz}
\usetikzlibrary{arrows.meta, positioning}
\usepackage{comment}
\usepackage{url}

\usepackage[utf8]{inputenc} 
\usepackage[T1]{fontenc}    
\usepackage{hyperref}       
\usepackage{url}            
\usepackage{booktabs}       
\usepackage{amsfonts}       
\usepackage{nicefrac}       
\usepackage{microtype}      
\usepackage{xcolor}         
\usepackage{authblk}

\usepackage{ifthen}

\newcommand{\shortfundingstatement}{
    SNL is managed and operated by NTESS under DOE NNSA contract DE-NA0003525

}

\newcommand{\mediumfundingstatement}{
    Sandia National Laboratories is a multimission laboratory managed and operated by National Technology \& Engineering Solutions of Sandia, LLC, a wholly owned subsidiary of Honeywell International Inc., for the U.S. Department of Energy’s National Nuclear Security Administration under contract DE-NA0003525. SAND\#0000-XXXXX

    This paper describes objective technical results and analysis. Any subjective views or opinions that might be expressed in the paper do not necessarily represent the views of the U.S. Department of Energy or the United States Government.
}

\newcommand{\longfundingstatement}{
     This article has been authored by an employee of National Technology \& Engineering Solutions of Sandia, LLC under Contract No. DE-NA0003525 with the U.S. Department of Energy (DOE). The employee owns all right, title and interest in and to the article and is solely responsible for its contents. The United States Government retains and the publisher, by accepting the article for publication, acknowledges that the United States Government retains a non-exclusive, paid-up, irrevocable, world-wide license to publish or reproduce the published form of this article or allow others to do so, for United States Government purposes. The DOE will provide public access to these results of federally sponsored research in accordance with the DOE Public Access Plan \href{https://www.energy.gov/downloads/doe-public-access-plan}{https://www.energy.gov/downloads/doe-public-access-plan}.
}

\newcommand{\insertfundingstatement}[1]{
    \ifthenelse{\equal{#1}{short}}{
        \shortfundingstatement
    }{
        \ifthenelse{\equal{#1}{medium}}{
            \mediumfundingstatement
        }{
            \ifthenelse{\equal{#1}{long}}{
                \longfundingstatement
            }{
                \textbf{Error: Invalid funding statement type. Use 'short', 'medium', or 'long'.}
            }
        }
    }
}

\title{Neuromorphic Simulation of \textit{Drosophila Melanogaster}\\Brain Connectome on Loihi 2}
\author{Felix Wang, Bradley H. Theilman, Fred Rothganger,\\William Severa, Craig M. Vineyard, and James B. Aimone}
\affil{Neural Exploration and Research Laboratory \\ Sandia National Laboratories \\ Albuquerque, NM 87185}
\date{}

\begin{document}

\maketitle

\begin{abstract}
We demonstrate the first-ever nontrivial, biologically realistic connectome simulated on neuromorphic computing hardware.
Specifically, we implement the whole-brain connectome of the adult \emph{Drosophila melanogaster} (fruit fly) from the FlyWire Consortium containing 140K neurons and 50M synapses on the Intel Loihi 2 neuromorphic platform. This task is particularly challenging due to the characteristic connectivity structure of biological networks.
Unlike artificial neural networks and most abstracted neural models, real biological circuits exhibit sparse, recurrent, and irregular connectivity that is poorly suited to conventional computing methods intended for dense linear algebra.
Though neuromorphic hardware is architecturally better suited to discrete event-based biological communication, mapping the connectivity structure to frontier systems still faces challenges from low-level hardware constraints, such as fan-in and fan-out memory limitations.
We describe solutions to these challenges that allow for the full FlyWire connectome to fit onto 12 Loihi 2 chips.
We statistically validate our implementation by comparing network behavior across multiple reference simulations.
Significantly, we achieve a neuromorphic implementation that is orders of magnitude faster than numerical simulations on conventional hardware, and we also find that performance advantages increase with sparser activity.
These results affirm that today's scalable neuromorphic platforms are capable of implementing and accelerating biologically realistic models---a key enabling technology for advancing neuro-inspired AI and computational neuroscience.
\end{abstract}


\section{Introduction}

The FlyWire connectome, which is the first full connectome of an adult \textit{Drosophila melanogaster} nervous system, represents a major advance in anatomically capturing the non-trivial complexity of the brain~\cite{dorkenwald2024neuronal, lin2024network,schlegel2024whole}. The availability of a complete circuit diagram is seen as a critical step towards describing neural computation at scale~\cite{denk2012structural, sporns2005human}. While this structural description of the connectome lacks the richness of biological neuronal dynamics and learning,  and is thus only a first step~\cite{bargmann2013connectome}, it is nonetheless sufficient to constrain full-scale simulations that are capable of exhibiting and predicting new behaviors \cite{shiu2024drosophila}. 

The ability to leverage brain-like hardware has long been seen as critical for at-scale simulation of neural circuits~\cite{hbpnmc2013}.  While it is the case that simple nervous systems exist at a scale easily handled by modern computing technologies (the FlyWire connectome contains fewer neurons and synapses than many production artificial neural networks), mammalian brains exist at a scale beyond most computing platforms. For instance, the human brain is estimated to contain $10^{15}$ synapses, which would equate to roughly $1$ petabyte of memory ($1$ million gigabytes) for just the synaptic weights alone. For this reason, achieving the scale of the human brain with even simplified models requires computational resources at the scale of some of the world's largest high-performance computing platforms~\cite{lu2024simulation}. The move to more brain-like complexity in neurons or synapses or the exploration of complex behaviors would easily push at-scale human brain simulations beyond the capacity of today's Exascale HPC systems~\cite{wang2024path}. 

Although not yet at human scale, today's neuromorphic hardware is beginning to reach scales that are notionally comparable to mammalian systems~\cite{kudithipudi2025neuromorphic}. While these systems still largely rely on digital logic to implement individual neurons, they can achieve considerable efficiency advantages from light-weight spiking communication and representation as a spatially distributed graph. For this reason, in principle they should be ideally suited for modeling biological neural circuits, particularly spiking models. Furthermore, unlike graphics processing units (GPUs), most neuromorphic systems do not suffer a penalty when simulating networks with complex connectivity. 

In this paper, we demonstrate a neuromorphic implementation of the FlyWire model described in~\cite{shiu2024drosophila} on the Intel Loihi 2 neuromorphic platform. While the FlyWire connectome may be relatively small compared to modern neuromorphic systems, its computational graph presents characteristics that are considerably different from the large-scale neural simulations that have been implemented on neuromorphic hardware to date~\cite{knight2016synapse, sharp2012power, imam2020rapid}. We describe methods for addressing these graph complexities in light of hardware constraints, and report the resulting hardware-accelerated performance advantages. Through this, we demonstrate the utility of neuromorphic hardware platforms for supporting simulations in neuro-inspired AI and computational neuroscience research.

\section{Background}

The concept of taking inspiration from the impressive computational properties of natural brains to develop algorithms as well as computer architectures such as neuromorphic hardware is straightforward~\cite{hassabis2017neuroai}. Advances in artificial neural networks (ANNs) have encompassed both structural and functional developments~\cite{resnet2016,attention2017,goodfellow2016dl,dlsurvey2021}. The ability to propagate a gradient over multiple layers has enabled the ability to learn deep neural networks. With the ability to construct deep networks, explorations into connectivity patterns across increasingly sophisticated network structures have included motifs such as recurrent connectivity via feedback loops as well as residual connections whereby neurons are connected not just in adjacent layers but across multiple layers. Furthermore, the fields of neural architecture search, evolutionary, and genetic algorithms have sought algorithmic means of exploring the structural topology of neural networks to yield performant solutions~\cite{nas2019survey,schuman2020eons,schuman2024hairball}. An underlying motivation to these endeavours has been an aspiration to discern what the right neural network structure is, in the absence of a blueprint from biological brains.

\subsection{Neural Connectomes}

In neuroscience, connectomes provide an annotated enumeration of the neurons and synapses that constitute a brain. Neuroscience has pursued acquiring full and partial connectomes since the realization that brains are networks of neurons connected through synapses. The density, scale, and complexity of natural brains makes acquiring connectomes an extraordinarily technical and labor intensive process. Nevertheless, there now exist multiple full connectomes and detailed partial connectomes from a variety of species. Notable examples include the \textit{Caenorhabditis elegans} (C. elegans) connectome, the first complete connectome with 302 neurons and roughly 6702 synapses~\cite{white1986structure, varshney2011celegans}. A much larger neuroscience model, the larval zebrafish has about 100K neurons and millions of synapses, although the exact number is not known~\cite{ahrens2013zebrafish}. There are also some cortical models from larger animals, such as mouse visual cortex as taken from the MICrONS project and Allen Institute~\cite{microns2025functional, AllenInstitute_v1, billeh2020systematic}. These models look at the organization of neurons within a specified columnar volume of cortex and consist of roughly 75K and 52K neurons, respectively.

At large scales, many neural circuit models are built from populations of roughly homogeneous neurons along with connection probabilities between these populations. While the mouse visual cortex model is simplified like this, it is still fairly complex overall with the model developed by the Allen Institute consisting of hundreds of unique neuron models and connectivity across 17 different cell classes. Population models like these for the human brain also exist, with a commonly implemented spiking network model of the cortical microcircuit containing about 80K neurons in 8 distinct populations and 300M synapses in a columnar organization~\cite{potjans2012micro}.

For this work, we look at the Drosophila \emph{melanogaster}, which has had its connectome imaged to have 140K neurons and 50M synapses~\cite{dorkenwald2024neuronal, schlegel2024whole}. Unlike the the population-based models the larger network structure may be used to inform heuristics, for the purposes of computation, it is much more appropriate to treat the FlyWire network model as a large flat graph with non-uniform properties. Mapping this biological connectivity to hardware can thus be thought of as a ``general'' case for a neuromorphic compiler.

\subsection{Brain-scale Simulations}

There has not been a lack of attempts at facilitating neural simulations. While there remains interest in biophysically detailed simulations~\cite{awile2022modernizing}; tracking the rise of ANNs, in recent decades many neural simulation approaches have prioritized the simulation of large populations of relatively simple neurons with powerful simulation tools such as BRIAN~\cite{goodman2009brian, stimberg2019brian}. As with ANNs, the growing emphasis on large-scale networks has led to looking at more powerful computing systems, including GPU-acceleration~\cite{knight2018gpus, knight2021larger} and high-performance computing (HPC)~\cite{diesmann2001nest, wang2024scaling}. Through these methods, extremely large-scale neural simulations have been performed (e.g.,~\cite{lu2024simulation}), though these models are by necessity greatly abstracted from real brains both for performance reasons and the lack of well-described connectomes. Indeed, a recent analysis suggests that achieving human-brain scales connectivity with even modest realism will require far beyond today's HPC capabilities~\cite{wang2024path} and likely necessitates the use of specialized neuromorphic hardware.

The ability to perform brain-scale neural simulations was an explicit goal of the Human Brain Project and led to the development of the SpiNNaker~\cite{furber2020spinnaker,furber2014spinnaker} and BrainScales~\cite{schemmel2010wafer,schmidt2023clean} efforts.  In principle, neuromorphic hardware presents a compromise between GPUs, which have many cores but strongly prefer dense homogeneous networks, and multi-core HPC simulations, which are more flexible for complex and heterogeneous networks but have relatively few cores.

Today, there are many approaches to neuromorphic hardware that can reach the scales of animal brains~\cite{kudithipudi2025neuromorphic}. While these systems vary in implementation details (e.g., SpiNNaker systems~\cite{furber2016large,gonzalez2024spinnaker2} are ARM-based with an optimized network-on-chip for spiking communication, IBM TrueNorth~\cite{merolla2014million} and Intel Loihi systems~\cite{davies2018loihi} have specialized digital neuron accelerators, and BrainScales uses subthreshold analog CMOS acceleration), all aspire to the aforementioned goal of balancing an extremely large number of simple processors that are well suited for neuron and synapse computations. These systems are generally programmable as a graph of neurons and synapses that is fully-expanded on the hardware at run-time, that is, the model is not dynamically accessed is the case on a GPU. Aligning a graphical description of a neuromorphic program (e.g.,~\cite{aimone2019composing}) with the specific neuromorphic core capacities and connectivity restrictions defines an embedding problem whose cost and feasibility determines the efficiency of a neuromorphic implementation. 

Although many of these systems have been explored with generic cortical networks or similar (e.g., balanced excitatory-inhibitory networks)~\cite{rhodes2019spinncolumn,rathi2023survey,schmidt2025brianscales,severa2025benchmarking} as well as ANNs ~\cite{davies2021advancing,shrestha2024efficient,abreu2025neuromorphic,vineyard2019low,kelber2020mapping,rostami2022prop,nazeer2024language,alom2018deep,schmitt2017neuromorphic}, prior to this work, it has remained an open question whether neuromorphic systems could efficiently implement full-scale networks with biologically-defined connectivity. While neuromorphic systems are generically programmable, the suitability for real networks is by no means guaranteed as the connectivity of real biological networks includes neurons with extreme fan-outs that can surpass the low-level connectivity restrictions in hardware, requiring more complex graph embeddings (and hardware-level optimizations) to be used. Further, to date it has been impossible to explicitly test this suitability due to the lack of full connectomes to validate.

\subsection{Neuromorphic Platforms}

While early neuromorphic platforms were developed with the notional goal of large-scale \textit{in silico} neuroscience studies~\cite{furber2014spinnaker, merolla2014million, schemmel2010wafer}, the demonstration of large-scale neural simulations was rather limited~\cite{knight2016synapse}. In part this is because this hardware largely pre-dated the neuroscience definition of large-scale models---until recently the only full brain connectome was the non-spiking 302-neuron \textit{C. Elegans}, but it also proved a difficult to align generic neural simulations with the necessarily hardware constraints of this hardware.  For example, in order to reach a million neurons per chip, the IBM TrueNorth design was effectively limited to an unrealistic fan-in of 256 inputs per neuron~\cite{akopyan2015truenorth}. For these reasons, neural simulations began to look primarily towards HPC platforms~\cite{diesmann2001nest, wang2024scaling} as well as GPUs~\cite{knight2018gpus, knight2021larger}. Notably, as with ANNs, GPUs proved preferable to neuromorphic approaches (and simulation) when models could be described with significant homogeneity~\cite{knight2018gpus}.

Due to CMOS scaling and architectural advances, more recent neuromorphic platforms, such as the Intel Loihi~\cite{davies2018loihi}and Loihi 2 platforms~\cite{loihi2techbrief}, the SpiNNaker 2 platform~\cite{gonzalez2024spinnaker2}, and BrainScales 2~\cite{schmidt2023clean}, have addressed many of the connectivity and local memory challenges that made the earlier generation of scalable neuromorphic hardware poorly suited for biological neural simulations. Nevertheless, in the absence of concrete reference connectomes, the largest biological neural models continue to rely heavily on abstracted models with structured connectivity that are ideally suited for conventional hardware~\cite{lu2024simulation}. As such it has remained unclear how neural simulations will scale when more sophisticated heterogeneity and circuit complexity is incorporated into models~\cite{wang2024path}.

\section{Methods}

\begin{figure}[h!]
    \centering
    \includegraphics[width=.8\linewidth]{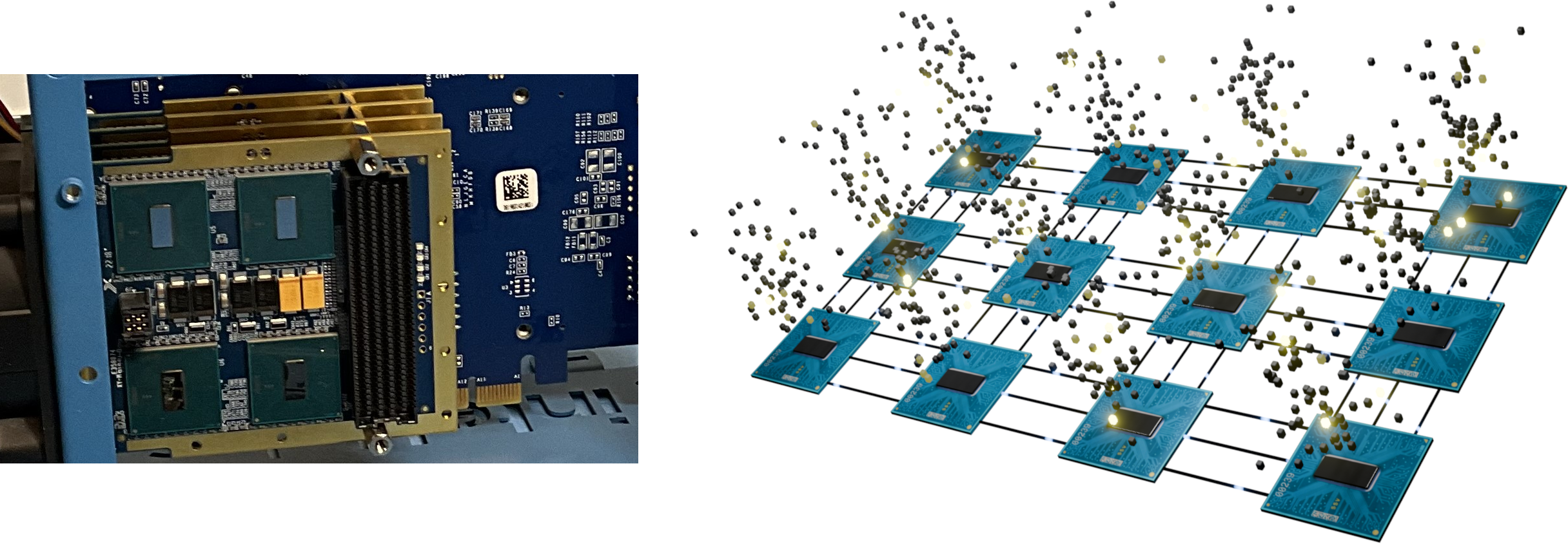} 
    \caption{Photograph of an Intel Loihi 2 Kapoho Point system (32 chips) (left), and a visualization of the spiking activity of the FlyWire network simulated on 12 chips (right).}
    \label{fig:method_loihi_hw}
\end{figure}

The central challenge for mapping the FlyWire network onto Loihi 2 is efficiently partitioning the irregular graph structure onto the highly distributed, memory-limited neurocores. Unlike on standard von Neumann architectures, where the details of the memory architecture are largely abstracted to one contiguous block in which only the total amount a program uses is relevant, this abstraction is not suitable for emerging neuromorphic platforms such as Loihi 2 in which processing is co-located with memory. In fact, within single chips the memory locality considerations more closely reflect what may be needed in high-performance computing (HPC) applications that deal with model parallelism when a model is too large to fit on any single node. On neuromorphic systems, it is useful to consider that the computing and memory resources of distributed nodes are much finer-grained than the HPC systems, with only a small amount of locally accessible memory available per neurocore.

At a high level, the process of converting the FlyWire network as simulated on a conventional simulator (Brian 2) onto a neuromorphic hardware accelerated implementation (Loihi 2) proceeds through a series of steps manipulating and reorganizing the relevant data structures as well as performing some appropriate computational approximations. To support this process, we used the STACS simulator~\cite{wang2015simulation,wang2024scaling} as an intermediate translation layer between the contiguous network structure and floating-point neuron dynamics of Brian 2 and the partitioned network structure and fixed-point neuron programs on Loihi 2.

Starting from the reference Brian 2 simulation, a corresponding simulation was first replicated in STACS. This intermediate reference simulation was used to both bring the FlyWire network into a partitioned intermediate representation through the SNN-dCSR format and also serve as a foundation for modifying the neuron dynamics through a more imperative way of representing the neuron models. These dynamics modifications addressed some of the practical limitations to the microcoded neuron programs on Loihi 2, including the effect of using fixed-point arithmetic, timestep discretization, weight quantization (and capping), and limited accumulation variables for input integration. The modified network in STACS could then be used as a more direct reference simulation for the hardware implementation.

Once in a partitioned data format, the FlyWire network is then repartitioned to better fit onto the computational resources across a set of Loihi 2 chips. This compilation step also involves computing some additional intermediate data structures more efficiently compress the connectivity information. Finally, these memory-efficient self-contained network partitions are configured onto the Loihi 2 system through its low-level NxCore API for execution. We walk through these steps in the following sections.

\subsection{FlyWire Network Model}

The network model for FlyWire we implemented is replicated from~\cite{shiu2024drosophila}. This is a simplified point-neuron model containing roughly 140K and 15M synapses. Compared to the more complete version of the model containing roughly 50M synapses, a simplification was performed to condense synapses with the same source and target neurons into the same connection variable (and scaling the weight appropriately). Furthermore, the neuron dynamics have been simplified to a two-state current-based leaky-integrate-and-fire neuron with refractory period. The model equations for the neuron are shown in Equation \ref{eq:neuronmodel}.

\begin{equation}
    \begin{split}
        \frac{dv}{dt} &= (v_0 - v + g) / \tau_{m} \quad \text{(unless refractory)}\\
        \frac{dg}{dt} &= -g / \tau_{g} \quad\quad\quad\quad\quad\; \text{(unless refractory)}\\
        \text{if }&\; v > v_{th} \quad \text{:} \quad v = v_{r}\\
        \quad&\;\quad\quad\quad\quad\quad\;\, g = 0\\
        \quad&\;\quad\quad\quad\quad\quad\;\, \text{enter refractory for } \tau_{ref}
    \end{split}
    \label{eq:neuronmodel}
\end{equation}

The time constant for the membrane potential, $\tau_m$, is set to 20ms, and the time constant for the conductance-like input, $\tau_g$, is set to 5ms. The refractory period, $\tau_{ref}$, for all neurons is set to 2.2ms (unless it is directly stimulated by a Poisson input). The values for the membrane voltage resting and reset potential, $v_0$ and $v_r$, respectively, are set to 0, and the threshold potential, $v_{th}$, is set to 7mV.

\begin{figure}[h!]
    \centering
    \includegraphics[width=.7\linewidth]{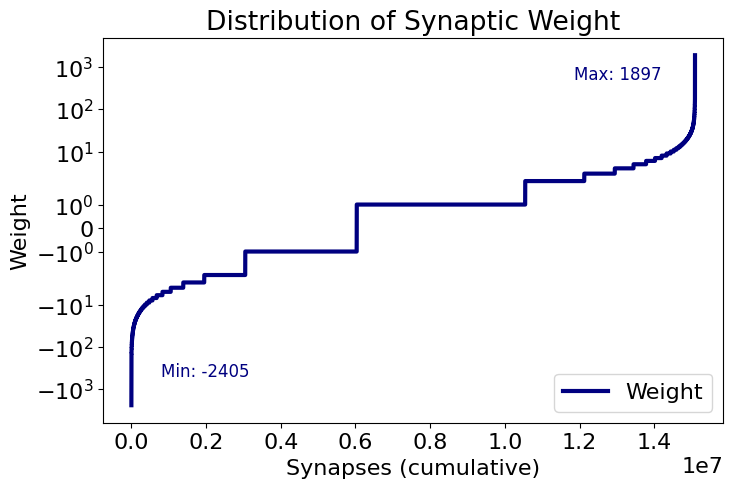} 
    \caption{Distribution of synaptic weight values across all synapses (cumulative sorted order for clarity), inhibitory weights are negative, excitatory weights are positive}
    \label{fig:weight_distribution}
\end{figure}

The synaptic weights between neurons are static and may be either excitatory (positive) or inhibitory (negative). These are quantized to integer numbers with a fairly large range between -2405 and 1897 due to outliers, but where the majority of the weights are modest with a magnitude under 100. This distribution of weights is shown in Figure \ref{fig:weight_distribution}. The weights are scaled by a value of 0.275mV prior to being added to the conductance-like state variable, $g$. The synaptic delay for all connections are set to the same constant of 1.8ms.

\begin{figure}[h!]
    \centering
    \includegraphics[width=0.7\linewidth]{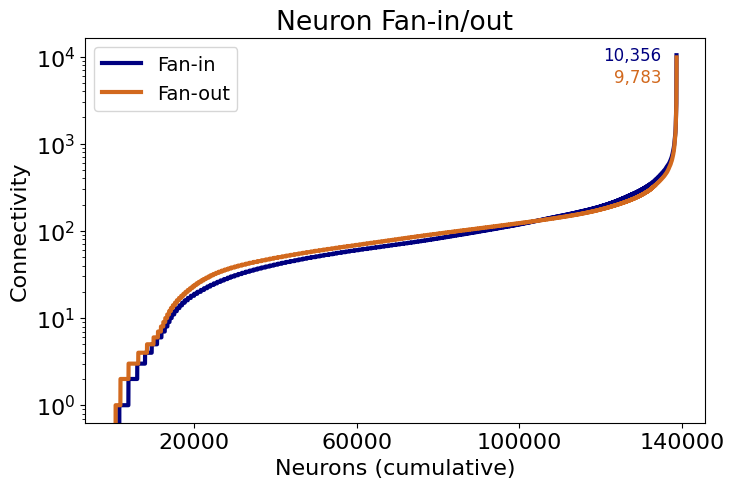}
    \caption{Distribution of fan-in and fan-out across all neurons (cumulative sorted order for clarity)}
    \label{fig:faninout}
\end{figure}

Despite these simplifications onto a point-neuron model, much like the distribution in weights, the connectivity structure in terms of the fan-in and fan-out between neurons also shares a similar structure where most neurons have a modest number of connections but several outliers have an order of magnitude higher connectivity. Here, the number of connections range from tens to tens of thousands with the max connectivity reaching 10,356 for fan-in, and 9,783 for fan-out. This distribution of connectivity per neuron is shown in Figure \ref{fig:faninout}. It is this biologically-realistic but uneven distribution in the network connectivity structure that makes many heuristic approaches for partitioning and mapping onto hardware that assume some form of regularity largely ineffective (e.g. distributing neurons evenly across neurocores).

\begin{figure}[h!]
    \centering
    \includegraphics[width=0.8\linewidth]{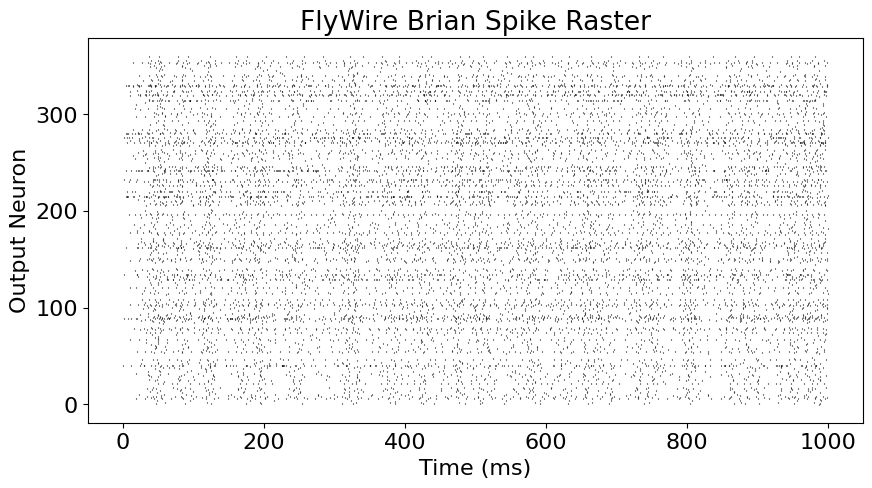}
    \caption{Spike raster of the sugar neuron experiment in Brian 2}
    \label{fig:raster_brian}
\end{figure}

In the effort to map this network onto Loihi 2, as a way to validate the correctness of our implementation and as a baseline of comparison, we furthermore implemented a simple sugar neuron experiment as taken from the same computational paper~\cite{shiu2024drosophila}. This experiment looked at the sensory pathways of the network model in response to the stimulation of a small subset of neurons (about 20) that were sensitive to the presence of sugar. These neurons were externally driven by Poisson inputs at 150Hz, and the spiking activity of the downstream neurons are recorded. What makes this experiment useful for our purposes is that the resulting spiking activity is fairly well contained to a few hundred neurons out of the whole network, making it easier to process and compare. Over a simulated duration of 1s, a typical spike raster from the Brian 2 simulator is shown in Figure \ref{fig:raster_brian}. Due to the random Poisson inputs, each simulation will result in slightly different spike times, but we can compare against the expected spike rates per neuron across multiple simulations.

\subsubsection{FlyWire in STACS}

To support the conversion process, we first reimplemented the FlyWire network model from Brian 2 directly in STACS. This allowed us to minimize the layers of translation between a simulated network model and its implementation in NxCore. We could also more easily modify the neuron model dynamics to match the various hardware execution models for validation. Working with the network connectivity information more directly also allowed for quicker iteration over measuring synaptic connectivity statistics and heuristics for partitioning onto the Loihi 2 compute and memory resources.

This step of moving between simulators is relatively straightforward as there are generally no significant compute resource constraints to worry about, and the bulk of the effort is simply defining the neuron models and marshaling the connectivity information into the appropriate data structures for instantiating the network.

For the neuron model, we took the system of ordinary differential equations as defined in Equation \ref{eq:neuronmodel} and developed a time-stepped model using forward Euler integration with an update interval of 0.1ms. STACS provides the ability to create custom user-defined models by extending a C++ base class that structures neuron dynamics as a set of mutable states and shared parameters. This allowed us to fully replicate the two-state LIF neuron model with refractory period that was used in Brian 2, as well as Poisson input neurons that targeted the conductance-like variable of the LIF neuron.

For the network connectivity, we used the Parquet formatted files provided by the FlyWire database, which lists all the synaptic connection pairs in a flattened table, and converted these into target-major adjacency lists to feed into the STACS build process, which allows for the instantiation of network information from user-defined build files. Synaptic weights and neuron states such as the refractory period were similarly instantiated through these build files. We used the same neuron ordering that was used in the Brian 2 simulation.

\begin{figure}
    \centering
    \includegraphics[width=0.8\linewidth]{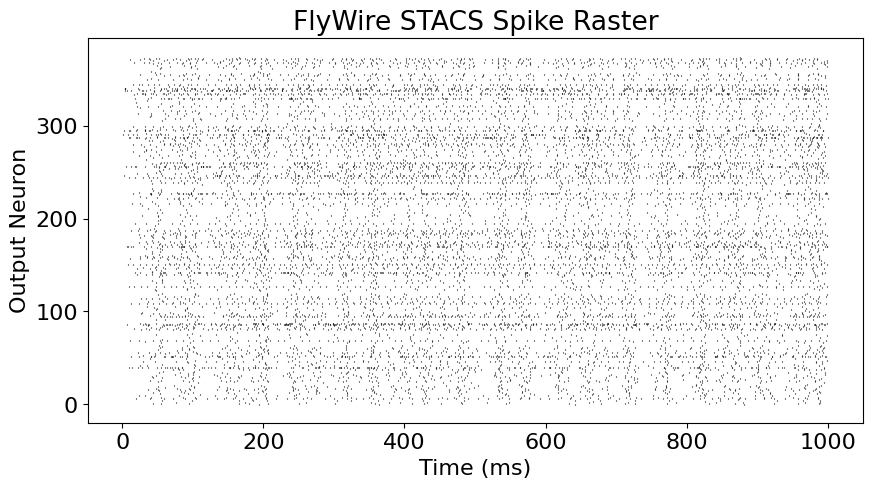}
    \caption{Spike raster of the sugar neuron experiment in STACS}
    \label{fig:raster_stacs}
\end{figure}

After building the FlyWire network in STACS, which generates partitioned network snapshot files in the SNN-dCSR format, the STACS simulation was run for 1s of simulated time. We parallelized the STACS simulation across 8 processes and used a communication time step of 1ms, which allowed for multiple computational updates to happen between needing to send spike information (the connection delay between neurons was 1.8ms). Spike output was collected for offline analysis and comparison to the Brian 2 ``ground truth''. A spike raster from the STACS simulation is shown in Figure \ref{fig:raster_stacs}.

\subsubsection{Comparing Spike Output}

Visually comparing the spike rasters from Brian 2 and STACS (Figures \ref{fig:raster_brian} and \ref{fig:raster_stacs}, respectively), we find that they look quite similar. However, it is challenging to determine just by eye how equivalent the outputs of these two simulations are. Furthermore, due to randomness from the Poisson inputs, there is also variability between simulations even from same simulator. 

\begin{figure}[h!]
    \centering
    \includegraphics[width=0.6\linewidth]{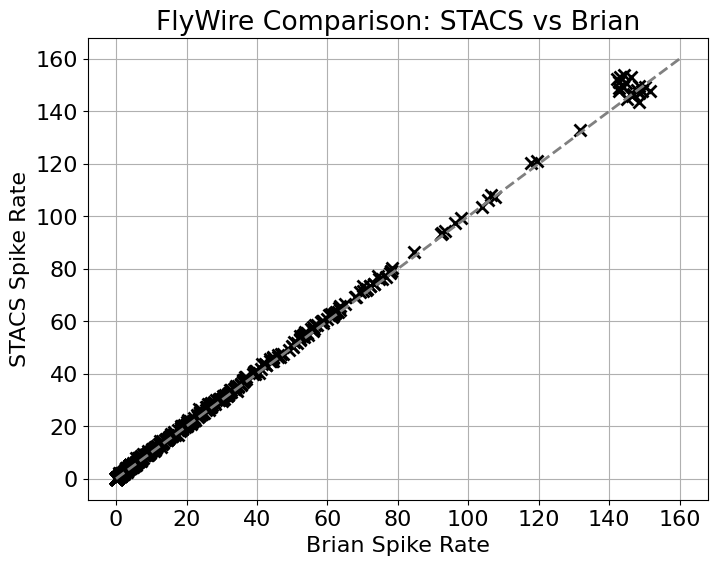}
    \caption{Comparing spike rates between STACS and Brian simulations, index-matched pairs of neurons between simulations marked with `$\times$', parity is shown as dashed line.}
    \label{fig:comp_stacs_brian}
\end{figure}

In order to validate our simulation results, as well as to use for future comparisons when executing the network on hardware, we used a relatively simple method that compared the computed spike rates between matched pairs of neurons against equality. Because we have the connectome data, the neuron index from one simulation method is matched with the neuron index of the other, and their average spike rates across multiple trials are plotted against each other. If the spike rates of the networks match, then they should fall along the parity line ($y=x$). Indeed, this is what we see across 10 trials for each simulator, as shown in Figure \ref{fig:comp_stacs_brian}.

\subsection{Mapping FlyWire to Loihi 2}

Equipped with a functional network model that supports custom repartitioning, flexible modifications to its dynamics, and a validation approach, we now turn our attention to the implementation details on Loihi 2. Unlike in simulation, moving to hardware involves a number of additional steps that take into account the computational resource constraints and format conventions. The main challenge is with respect to mapping the wide range of fan-in and fan-out in the biological network connectivity onto the more structured and limited local memory, with each of the distributed neurocores having access to 128KB synaptic memory. This memory is populated with data structures for the weights, delays, and indexes for where an incoming spike should be delivered. At the other end, the axon routing out of the neurons on a neurocore is also stored with information about the destination neurocore and the receiving axon index.

An important detail in the conversion process is that instead of storing and routing information based on global neuron indexes, the communication between any two given neurocores is coordinated using these axon indexes. This intermediate addressing system allows for some better memory efficiency, where each neurocore only needs to store information relevant to its local communication neighborhood. However, it also means that some additional bookkeeping must be performed to transform a unified connectivity adjacency matrix to a set of partitioned data structures coherent with this additional level of indirection. That is, instead of a spike being sent directly from a source neuron on one neurocore to a target neuron on another, it is set to a locally indexed synaptic memory data structure on the target neurocore which then delivers the spike locally. For clarity, we will describe this communication process from the sending side as axon routing, the addressing through the axon index, and the receiving side as synaptic delivery.

In addition to the synaptic connectivity, a few other key tasks involve setting up the Poisson input neurons, replicating the LIF neuron model dynamics, and collecting the spike output for comparison. For the Poisson inputs and LIF neuron models, Loihi 2 allows for the development of custom model dynamics through the use of microcoded programs. These allow for user-defined variables, operate on fixed-point arithmetic, and have access to a local random number generator. For the spike output, Loihi 2 provides a limited amount of spike counters on its embedded CPU cores where neurons may additionally send spikes to be recorded. We describe our solutions to these tasks in more detail in the following sections.

\subsubsection{Neuron Models}

We approximated the LIF neuron model dynamics defined in Equation \ref{eq:neuronmodel} in Loihi 2 microcode. The microcode program for a neuron in Loihi 2 is responsible reading inputs, updating local state, and generating spikes. More specifically, inputs from receiving synapses are delivered into a set of dendritic accumulators arranged as a shift buffer, the accumulator value for the current timestep is processed by the neuron program and spikes locally execute an axon program that controls the axon routing out of the neurocore.

For our implementation, we carried over the simplified forward Euler integration method that was introduced in STACS model dynamics. The variables for membrane voltage, $v$, and the conductance-like input, $g$, were represented as fixed-point numbers and the parameters for updating these variables against the resting potential as well as the weight scaling for incoming spikes were passed as user-defined constants to the microcoded neuron program.

Loihi 2 executes according to discrete timesteps, to which we used a timestep counter for managing the refractory period, $\tau_{ref}$. We used both a simulated timestep of 0.1ms and 1ms (for faster execution). For the 0.1ms case, the values for the refractory period and synaptic delay were 22 timesteps and 18 timesteps, respectively. For the 1ms case, we approximated these both to 2 timesteps.

For the Poisson input neuron model, we adapted the parameters of a LIF neuron model with probabilistic spiking such that it would evaluate the probability of spiking each timestep. These used the LFSR random number generator provided per Loihi neurocore. The Poison inputs spiked at a rate of 150Hz and were assigned one per sugar neuron.

One of the more significant approximations to the neuron model dynamics was an change to how the Poisson inputs stimulated the sugar neurons. In the original Brian 2 model, the Poisson inputs would directly update the membrane voltage variable of the LIF neuron it targeted (forcing it to spike). While this behavior is technically possible on Loihi 2 by allocating multiple dendritic accumulators per neuron, this would also greatly increase the complexity of the microcoded neuron model as well as the synaptic indexing. For our implementation, we opted to stay with a simpler input method where incoming spikes would only target and update the conductance-like input, which was the behavior between neurons of the main FlyWire network model.

\subsubsection{Synaptic Connectivity}

Putting aside the irregular connectivity structure, the individual synaptic connections between neurons of the FlyWire network are actually quite simple. Each connection has a constant weight value which is added to the conductance-like input variable of the target neuron. There is no learning, and all of the connections in the network also share the same 1.8ms delay. These weights and delays were stored in the synaptic memory alongside a local index for the dendritic accumulator an incoming spike should update on the receiving neurocore. The main complexity in the synaptic connectivity comes from the data structures involved in spike routing.

With respect to its representation in a more conventional SNN simulator like Brian 2, the synaptic connections in a network model may all be collected into a single global data structure such as a sparse weight matrix, and the propagation of spikes could in theory be performed through matrix multiplication with a vector of spiking neuron indices. Of course, this naive description of spike propagation is computationally wasteful when the spiking activity is sparse, and in practice there are many methods to take advantage of sparsity to accelerate the communication workload, even on GPU~\cite{knight2018gpus}.

On a neuromorphic platform like Loihi 2, there is no global memory space, and the connectivity information must be distributed across multiple partitioned data structures. This is analogous to the problem of domain decomposition in HPC where a large model may be distributed across multiple compute nodes, only at a much finer-grained scale. In more regular applications, such as a stenciled physics simulation, the exchange boundaries between compute nodes are referred to as ghost or halo regions, and this neighboring data must be kept up-to-date in order for the simulation to proceed correctly. We may extend this concept to reason about the synaptic connectivity, with the main differences that the exchange neighborhood is now determined by the edge cut between graph partitions and the spiking information that keeps the data dependencies up-to-date is sparse.

For a given partition of neurons that are assigned to a neurocore, the collective fan-in and fan-out of its source and target neurons, respectively, make up this exchange neighborhood. This locally relevant information enables a partition to be computationally self-contained except for the exchange of spike events. Instead of routing spikes based on global neuron indexes, Loihi 2 uses axon indexes that are local per neurocore, and employs a destination-based routing scheme, where a spike message includes the target chip, core, and axon index. This allows for both a greater degree of memory efficiency through more compact index spaces and a lower computational overhead by omitting routing table lookups.

However, this also requires that this more compact routing information be pre-computed, which must take into account the specific partitioning and mapping of neurons to neurocores which affects the addressing. This conversion from global and unified to local and distributed communication information can be a challenging task for neuromorphic compilation onto Loihi 2 and in general, especially when a valid partitioning solution must be iteratively computed.

To support this conversion process, we leveraged the partitioned data structures implemented in STACS. Given a computed partitioning, these provide a set of compact adjacency lists that contain the global indexes of the source and target neurons to and from a given partition, respectively, as well as a supporting list of the cumulative distribution of neurons per partition. Importantly, these global neuron indexes reflect a sequential ordering with respect to the partition order, making it straightforward to compute which partition a neuron is located from its index. When a network is repartitioned in STACS and the neurons are migrated to an updated partition ordering, these global neuron indexes are recomputed and reassigned.

In effect, this progresses the conversion process from simulation to hardware into an intermediate stage where the communication information is global and distributed. From here, it becomes much more straightforward to compute the local information. Given a mapping of partitions to neurocores, the chip and core index of a target neuron can be found by matching its index against the cumulative neuron distribution list. We can also find the subset of incoming and outgoing connections shared between a pair of partitions by matching against their respective global index ranges. While this information alone is sufficient for a naive point-to-point assignment of axon indexes between neurocores, the added level of indirection also allows for more compressed communication schemes.

\subsubsection{Communication Schemes}

We may take advantage of the addressing system that Loihi 2 uses, where spikes are routed through local axon indexes, to make the communication between neurocores more efficient. Instead of providing information about the source or target neuron index in the spike message explicitly, the information about where an incoming spike to a neurocore may have been sent from and where that spike should be locally delivered are implicitly provided by the axon index. What makes this useful for compression is that what an axon index represents can be overloaded.

A fairly straightforward way to use this for compression is to use the axon index to deliver a single incoming spike to multiple target neurons on the neurocore. Here, each of the axon indexes on the neurocore as well as their corresponding synaptic delivery lists would be assigned to each of the unique incoming neurons. The main advantage for this method is reducing the volume of spikes that must be communicated through the network. When a source neuron spikes on a sending neurocore, it only has to send a single spike message containing its unique incoming axon index to each of the receiving neurocores where it has target neurons. This reduces the effective fan-out of the sending neurons.


This method is similar to how a parallel simulator such as STACS may aggregate event messages from one process to be routed to the others (e.g. through multi-cast). It works especially well under the condition where there may be many targets on a given partition. However, one of the key drawbacks of this method is that the total amount of synaptic memory in terms of weights, delays, and local neuron indexes still has to be stored (even if organized jointly according to shared inputs). On a neuromorphic system such as Loihi 2 where memory per neurocore is limited, this compression scheme still led to the need to cap the fan-in of the outlier neurons (for our experiments, this was set to 4096).

Although somewhat counter-intuitive, another way to compress the connectivity information is to identify redundancies in the synaptic memory and allow the axon index to be shared across multiple sources. This is possible for the FlyWire network because the weights are static, rather than being uniquely updated from local learning rules. Here, we took advantage of the spike processing behavior of Loihi 2 where incoming spikes are buffered and their synaptic delivery is executed independently, allowing us to overload the axon index from the sender side. To identify redundancies, we revisit the distribution of weights over the network (Figure \ref{fig:weight_distribution}) and note that a significant fraction of them are just $\pm 1$ (before being multiplied by the weight scaling factor). Furthermore, if we consider the common practice of quantizing weights, we realize that multiple source neurons will likely have the same effect on the target neuron. Or more precisely for Loihi 2, the same effect on the dendritic accumulator when also taking into account the unique weight and delay pairs.

In our case, we used fixed-point integer weights with a bit width of 9 (one for the sign), and capped the outlier weights above and below to 255 and -256, respectively. All of the delays were the same. Compressing the connectivity information in this way meant that for any given neuron, its effective fan-in could only ever be as large as its weight quantization, and our theoretical max of 512 was significantly lower than the original max fan-in of 10,356. In practice, the total number of unique weights per neuron was even lower.

Aside from mainly being suitable for network `inference', where weights are assumed to remain static, one of the key drawbacks to this method of sharing sources is that a neuron would have to send its full fan-out volume of spike messages when it spikes. Such a communication strategy would likely be prohibitively expensive if it were performed as described within the context of a spiking simulator on conventional hardware, where the standard practice would be to aggregate messages and communicate information in as few exchanges as possible. However, an important trade-off here is that the Loihi 2 hardware has been designed to optimize the transmission of these smaller spike messages, but is far more limited on memory.

\begin{figure}[t!]
    \centering
    \includegraphics[width=0.8\linewidth]{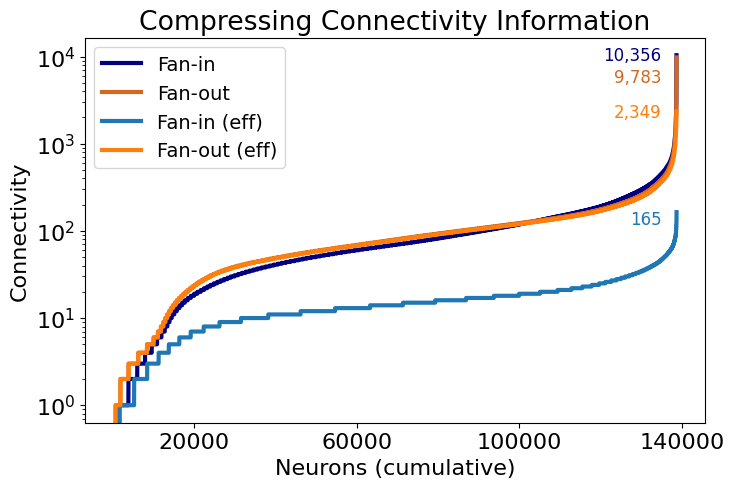}
    \caption{Effective Fan-in/out based on the two different compression schemes}
    \label{fig:connpress}
\end{figure}

We may refer to these two compression methods around the axon index as shared synaptic delivery (where an axon index is used for multiple targets) or shared axon routing (where an axon index is used for multiple sources). These reduce the effective memory required for storing the fan-out and fan-in information, respectively. Ideally, there could be more complex compression methods that could strike a balance between the amount of spikes and synaptic memory needed. We plot the effective fan-in and fan-out under these two compression schemes in Figure \ref{fig:connpress}. For shared synaptic delivery, the effective fan-out per source neuron is actually dependent on the resulting distribution of its target neurons to neurocores, and we use the values from a valid partitioning discussed more in detail in the next section. For shared axon routing, the effective fan-in per target neuron is independent of the partitioning.

What we find is that while both compression methods greatly reduce the amount of effective fan-in or fan-out, the shared axon routing is an order of magnitude more efficient, going from a max original fan-in of 10,356 to only 165. This meant that unlike the shared synaptic delivery method, we also did not need to artificially cap the fan-in of the outlier neurons due to synaptic memory limitations and could map the full FlyWire connectome on to Loihi 2. Furthermore, by more efficiently utilizing the synaptic memory, we could also fit more neurons per neurocore through this approach. We unfortunately still ran into some limitations on the maximum size of the axon programs controlling the axon routing, which led to the neurocores being underutilized.

\subsubsection{Partitioning to Neurocores}

Now that we have the microcoded neuron models and the partitioned data structures for representing the synaptic connectivity on Loihi 2, what remains is to compute a valid partitioning of neurons to neurocores that satisfies hardware limitations. Although we leveraged the network repartitioning functionality of STACS, the default graph partitioning scheme that is employed splits neurons evenly across partitions. While this is useful for load-balancing computation on a conventional system, it does not take into consideration the irregular distribution of synapse connections and their effect on memory utilization. To address this, we implemented a relatively simple partitioning scheme to better balance around the memory constraints, and supplied the resulting partition ordering to the repartitioning utility in STACS.

While common practice, the main issue with placing an even number of neurons per neurocore is that it will either overcommit or undercommit the available memory on the neurocores for networks such as FlyWire. If the number of neurons is set too high, the neurocores containing outlier neurons will exceed the memory requirements, if the number of neurons is set lower to accommodate for these outlier neurons, then the majority of neurocores without outlier neurons will be severely underutilized.  Instead of distributing neurons solely based on neuron count then, we also want to consider the synapse count of these neurons. With respect to the communication compression strategies, this can also be extended to capture the effective counts for fan-in and fan-out per neuron.

To map the FlyWire network onto Loihi 2, the custom partitioning scheme we implemented was a simple greedy partitioning that distributes neurons based on the estimated accumulated synaptic memory used for axon routing and synaptic delivery. For each partition, we set a capacity condition for the number of neurons and the number of incoming and outgoing connections. Neurons were assigned in ascending indexical order to the list of available partitions. If the neuron assignment to a partition would exceed its capacity in any of the three conditions, the neuron would be assigned to the next available partition (in ascending indexical order). After assignment, if the remaining capacity of a partition in any of the three conditions were sufficiently exhausted, then the partition would be marked as full and no longer receive any new assignments. This procedure continued until all of the neurons were assigned.

There were a few other memory limitations that we also ran into when executing the resulting partitioned network on Loihi 2. For the shared synaptic delivery, because the synaptic memory was also used for storing the axon routing programs and the incoming spike buffer, we had to adjust our incoming connection capacity to leave some more space available. Furthermore, due to the number of bits required to store weights, delays, and indexes, it was actually impossible to fit some of the outlier neurons with high fan-in on any of the neurocores altogether. This led us to artificially limit the fan-in connections of these neurons to 4096 with a combination of sampling and weight rescaling. For the shared axon routing, there was actually an additional memory limitation in the maximum size of the axon programs controlling the axon routing out of a neurocore. Because this compression method was so effective at reducing the memory required for synaptic delivery, this additional limitation actually left many of the neurocores underutilized. However, unlike the shared synaptic delivery approach, it was able to fit the entire FlyWire connectome onto Loihi 2 without any structural changes.

Taking into account these additional hardware limitations, we also adjusted the parameters of our partitioning scheme such that the resulting partitioning would utilize whole chips of Loihi 2, where each chip contained 120 available neurocores. The resulting partitioned network was mapped one partition per neurocore through the NxCore API, filling in the local data structures according to the chosen communication scheme. Using shared synaptic delivery, this resulted in mapping a slightly modified FlyWire connectome onto 20 chips (2400 neurocores). In estimating the memory utilization for partitioning, although the size of the axon routing programs were technically dependent on the number of partitions, its impact was mostly negligible compared to the synaptic memory used for the weights, delays, and indexes. More successfully, using shared axon routing allowed us to map the full unmodified FlyWire connectome onto 12 chips (1440 neurocores), resulting in a much better utilization of resources. Furthermore, it was easier to estimate the impact of this communication scheme on our neurocore capacity conditions because compression of the synaptic delivery information only required local fan-in information.

\begin{figure}[h!]
    \centering
    \includegraphics[width=0.75\linewidth]{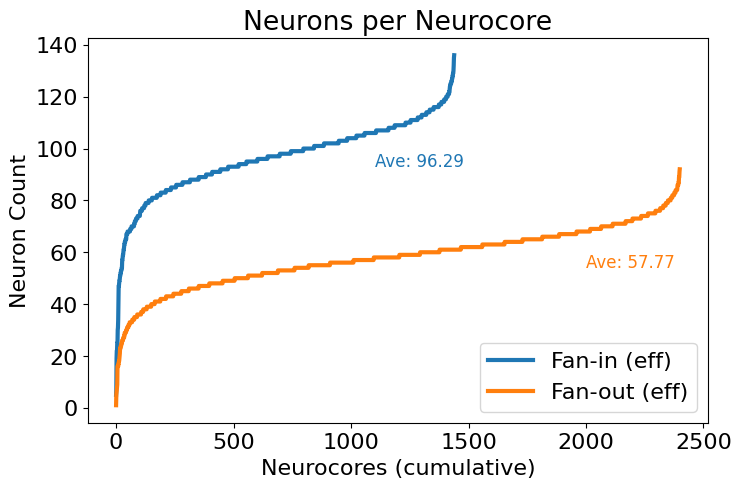}
    \caption{Number of neurons per neurocore (cumulative sorted order for clarity)}
    \label{fig:neuronspercore}
\end{figure}

We may look at the memory utilization for these two resulting partitioning outputs. The distribution of neuron count per neurocore is shown in Figure \ref{fig:neuronspercore}, and we see that instead of a even distribution, there are some outlier neurocores that have much fewer neurons as well as a few neurocores that have relatively more neurons. Although both compression methods result in roughly the same distribution pattern, we also see that shared axon routing results in a higher density of neurons per neurocore from fitting them to fewer partitions.

\begin{figure}[h!]
    \centering
    \includegraphics[width=0.75\linewidth]{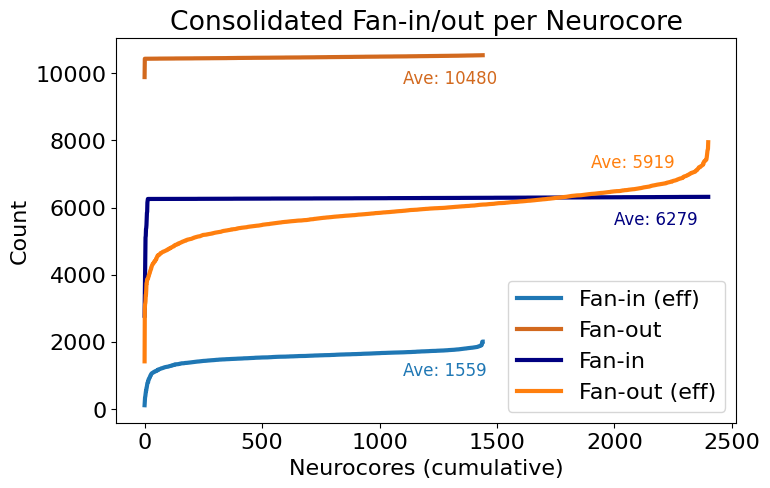}
    \caption{Number of fan-in/out connections per neurocore (cumulative sorted order for clarity)}
    \label{fig:synapsespercore}
\end{figure}

The distribution of fan-in and fan-out per neurocore (as accumulated over their neurons) is shown in Figure \ref{fig:synapsespercore}, where we immediately notice that some of the lines are relatively flat. These indicate the main limiting quantity in the partitioning scheme for a chosen compression method. For shared synaptic delivery (fan-in and fan-out (eff) lines), the synaptic memory was primarily limited on the storage of fan-in connections. For shared axon routing (fan-in (eff) and fan-out lines), the main limitation was actually not the synaptic memory but rather the maximum size of the axon programs controlling the axon routing.

\begin{figure}[h!]
    \centering
    \includegraphics[width=0.75\linewidth]{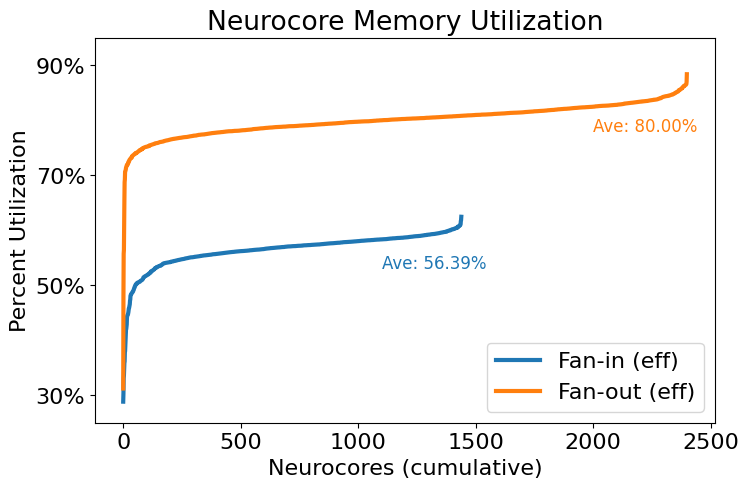}
    \caption{Neurocore memory utilization (cumulative sorted order for clarity)}
    \label{fig:memutilization}
\end{figure}

Finally, the measured memory utilization per neurocore is shown in Figure \ref{fig:memutilization}. While there are some outlier neurocores with less memory utilization than others, we mostly see that the memory utilization per neurocore is relatively uniform otherwise. Our partitioning scheme did not include any additional fine-tuning steps to redistribute neurons after an initial partitioning.

We found that despite the shared axon routing communication scheme resulting in far fewer partitions, its average memory utilization (56.39\%) per neurocore was also significantly less than that for shared synaptic delivery (80.00\%). While we have not investigated how useful the shared axon routing approach is in general, especially with its trade-off in increased spike volume, at least for the FlyWire network, it was instrumental to addressing the challenges of memory utilization on Loihi 2. From some preliminary experiments, we also found that the network partitioned through shared axon routing was also faster to execute than the network partitioned through shared synaptic delivery. For these reasons, our remaining comparison experiments were carried out using the shared axon routing communication scheme.

\subsubsection{Collecting Output}

\begin{figure}[h!]
    \centering
    \includegraphics[width=0.8\linewidth]{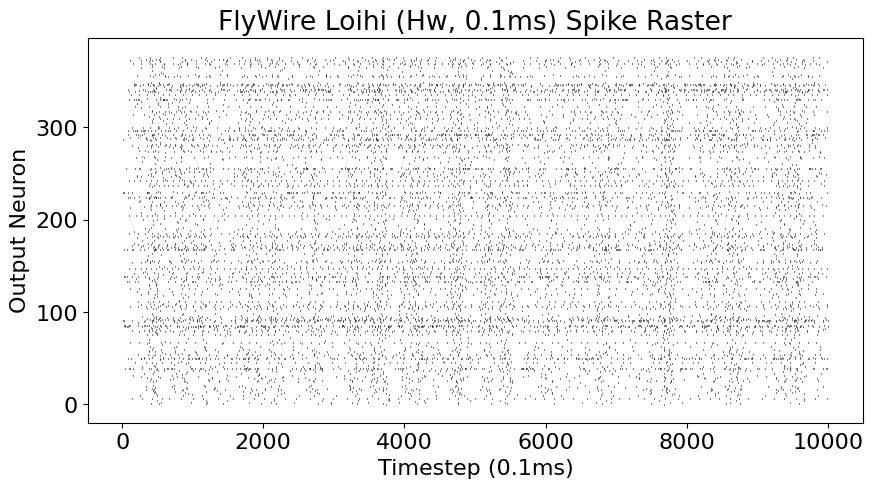}
    \caption{Spike raster of the sugar neuron experiment on Loihi with 0.1 ms timesteps}
    \label{fig:raster_loihi}
\end{figure}

To extract the spiking activity of the network for validation, we used the spike counters provided by the NxCore API. These were implemented on the embedded CPU cores per Loihi 2 chip, and there are a total of 992 spike counters available, with 224 of these additionally able to record payload information sent alongside the spike message. Because the number of neurons on a Loihi 2 chip (about 11.5K on average) would otherwise exceed the total number of spike counters, we used the payload functionality to overcommit the available spike counters using the neuron index as a payload. In our case, we utilized one payload-based spike counter per neurocore for a total of 1440 total spike counters across 12 chips. Through this, we were able to generate a spike raster from our replicated sugar neuron experiment on Loihi 2 as shown in Figure \ref{fig:raster_loihi}. Comparing this spike raster visually to the one from Brian 2 and STACS (Figures \ref{fig:raster_brian} and \ref{fig:raster_stacs}), we see that they are indeed quite similar.

There were also some disadvantages that came with using these spike counters, however. One of the unfortunate behaviors of the payload-based spike counter was that if multiple spike messages to the same counter were received in the same timestep, although it updated the total count to match the number of spikes, only the most recent payload would be recorded. This meant that occasionally by random chance the neuron index of a spike would be missed. By cross referencing against the total spike count, this fraction of missing spike data for the sugar neuron experiment was fortunately very low at less than 0.1\%, and by collecting spiking data over multiple runs we could still adequately perform our statistical validation comparing average spike rates. Perhaps the biggest drawback to using the spike counters is that because they require synchronized communication with the embedded CPU, it significantly slows down the execution of the network compared to if the spike messages were kept solely within the neurocore interconnects. There are some more recent methods that allow for non-blocking communication methods to send spike messages through an Ethernet connection directly to the host machine, but this functionality is still relatively new and not available for general use at the time of writing. Thus for our experiments, we used the spike counters when we needed to measure spike rates for validation, and omitted their use when we were measuring performance.

\subsection{Simple Scaling Study}

In addition to replicating some the sugar neuron experiment on Loihi 2, we also performed a fairly simple scaling study with respect to network activity. As can be seen in the spike rasters, the total number of active neurons were only a small fraction of the network, about 0.3\% of the 138K total neurons. While the average spike rate across just these active neurons was about 30Hz, effective spike rate across the whole network was only about 0.1Hz. This was because for this experiment, there was no baseline background activity and the only input to the network came from about 20 Poisson spiking input neurons. While this smaller scale experiment was useful for validation purposes, we also wanted to look at how performance scaled with increased spiking activity.

To provide a degree of uniformity, we looked at simply setting a baseline background spiking rate for all of the neurons and allowing their spikes to propagate through the network. For consistency, and to prevent potential runaway excitation, we set the weight scaling factor for the synaptic inputs such that the effects of the incoming spikes would be negligible in determining whether the target neuron spiked. Instead, we added probabilistic spiking functionality to the microcoded neuron model and parameter to control its spiking rate, similar to the mechanism from the Poisson input neurons. This way, we could execute the same FlyWire computational workload in terms of neuron model updates and synaptic communication, except for the spiking activity being driven by the baseline background spiking rate. We scaled these spike rates from 0.5Hz up to 40Hz, roughly corresponding to sparse to denser spiking activity throughout the network.

\section{Results}

\begin{figure}[h!]
    \centering
    \includegraphics[width=0.6\linewidth]{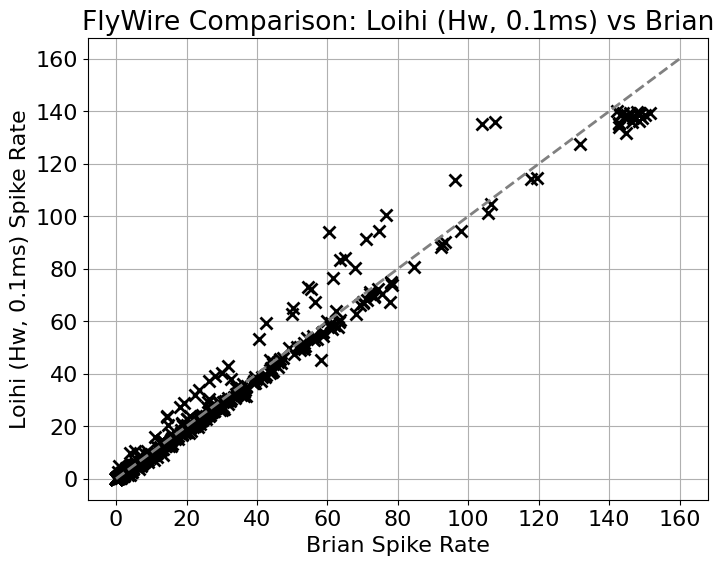}
    \caption{Comparing spike rates between Loihi 2 neuromorphic hardware implementation and Brian 2 simulation, index-matched pairs of neurons are marked with `$\times$', parity is shown as dashed line.}
    \label{fig:nxcorep1ms_brian}
\end{figure}

We compared the spike outputs from our Loihi 2 implementation to the reference Brian 2 simulation by matching the average spike rates between neurons by index. This allowed us to identify potential differences from our approximation that may not be visually apparent from the spike raster. We collected spike rates from ten independent trials and generate the parity plot, shown in Figure \ref{fig:nxcorep1ms_brian}.

We found that while the spike behavior from our Loihi 2 implementation largely followed the same pattern from Brian 2, there are also some noticeable differences between the two models. There are a set of neurons from Loihi 2 that follow a slightly off-parity line with higher than baseline average spike rates, and the cluster of neurons at the top-right side of the parity plot show lower than baseline average spike rates. 

\subsection{Sources of Model Differences}

Because of the model approximations we employed for our Loihi 2 implementation, we did not expect to see an exact correspondence with respect to the reference implementations in Brian 2 or STACS. However, we also wanted to validate if these approximations were sufficient to fully explain the differences between the two implementations. To support this analysis, we modified our intermediate STACS implementation to compare the effects of the model approximations independently and jointly on the resulting spiking behavior.

\begin{figure}[h!]
    \centering
    \includegraphics[width=0.45\linewidth]{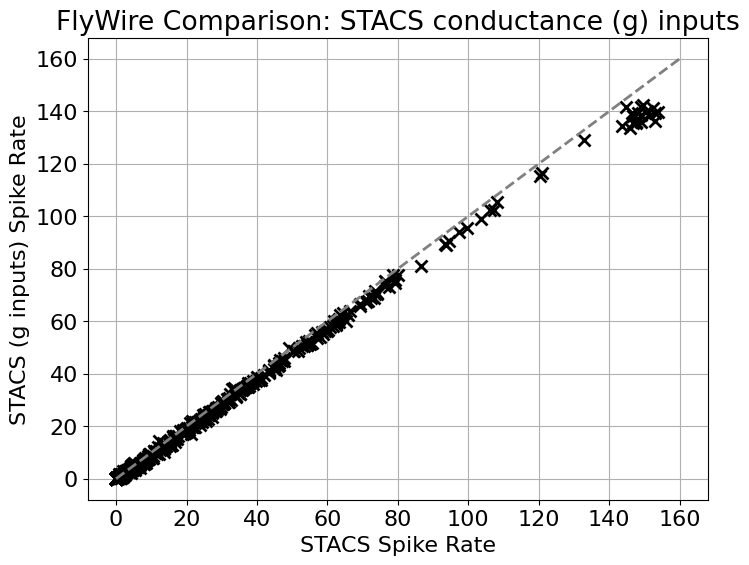}
    \includegraphics[width=0.435\linewidth]{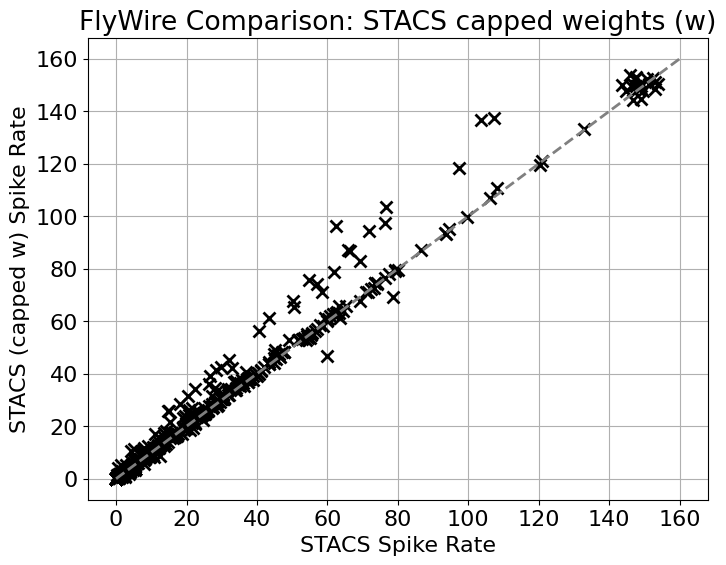} 
    \caption{The effect on spike rates for conductance-only inputs (left) and capped weights (right), in simulation}
    \label{fig:gonly_capped}
\end{figure}

One of these model approximations was how neurons only received spike inputs through its conductance-like variable, $g$, as opposed to the Poisson inputs of the Brian 2 model directly updating the membrane potential, $v$. This would lead to an integration delay to the membrane potential and also potentially alias consecutive inputs that happened to be close in timing. Another approximation was capping the integer synaptic weights to between 255 and -256 due to a bit width of 9 (with one bit representing the sign). While this captured the majority of weights in the network (see Figure \ref{fig:weight_distribution} for reference), this still resulted in 454 negative and 637 positive weights being capped (accounting for only 0.007\% of the total weights in the network). The parity plots between the STACS model incorporating these approximations and the baseline model is shown in Figure \ref{fig:gonly_capped}.

Looking at these parity plots, we notice that the individual effects correspond to the differences we identified in the spike behavior of the Loihi 2 hardware implementation in Figure \ref{fig:nxcorep1ms_brian}. For the approximation of conductance-only inputs, we see that the cluster of neurons on the top-right side of the parity plot are lower than the baseline average spike rates. We also notice a slight effect of spiking behavior slightly below parity for the remainder of the neurons in the modified simulation. For the approximation of capped weights, we notice that the cluster of neurons on the top-right are spiking normally, but that the slightly off-parity line of neurons with higher than baseline average spike rate is present. Incidentally, if we look at the individual pairs, they also look quite similar to the pattern produced by the Loihi 2 hardware implementation.

\begin{figure}[h!]
    \centering
    \includegraphics[width=0.45\linewidth]{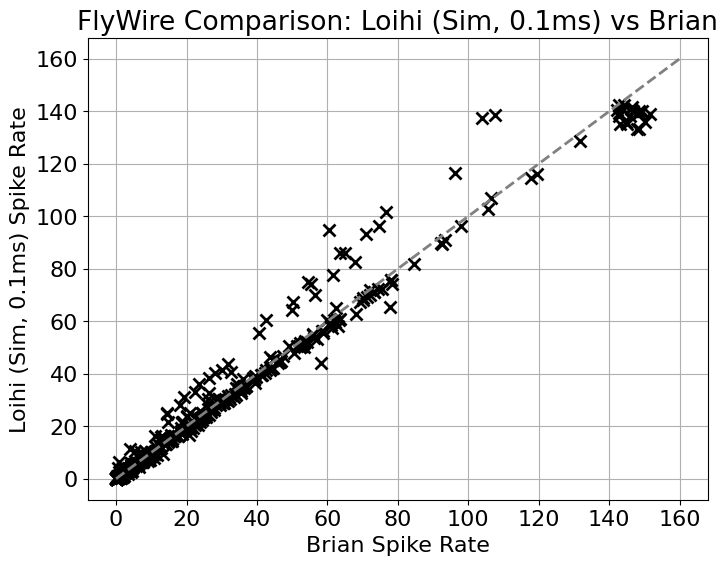}
    \includegraphics[width=0.445\linewidth]{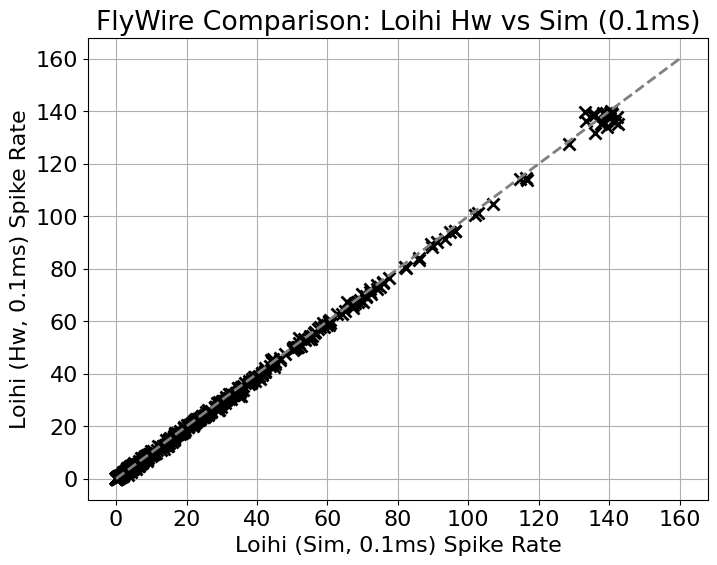} 
    \caption{Comparing the STACS behavioral simulation of our Loihi 2 hardware implementation to the Brian 2 reference simulation (left) and our actual Loihi 2 hardware implementation (right)}
    \label{fig:stacs_intermediate}
\end{figure}

Jointly combining the modifications from these approximations together, we essentially obtain a STACS behavioral simulation of our Loihi 2 hardware implementation. We can compare this version of the FlyWire model to both the Brian 2 reference simulation and the Loihi 2 hardware implementation. The parity plots for these comparisons are shown in Figure \ref{fig:stacs_intermediate}. We see a qualitatively similar parity plot from this Loihi 2 behavioral simulation to the Brian 2 reference simulation as we saw from Loihi 2 hardware (see Figure \ref{fig:nxcorep1ms_brian}). Furthermore, when comparing the average spike rates from our behavioral simulation to the actual hardware implementation, we obtain an equivalent parity plot. Consequently, we have suitably explained the differences in spike behavior between simulation and hardware.

\begin{figure}[h!]
    \centering
    \includegraphics[width=0.45\linewidth]{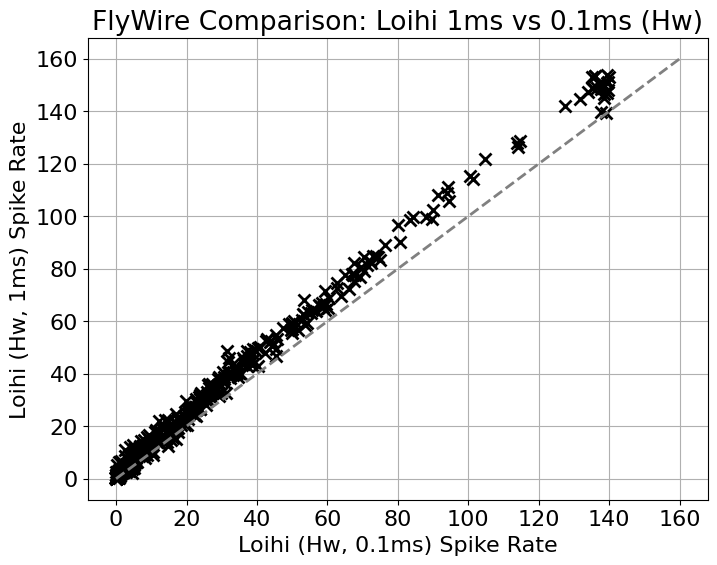}
    \includegraphics[width=0.45\linewidth]{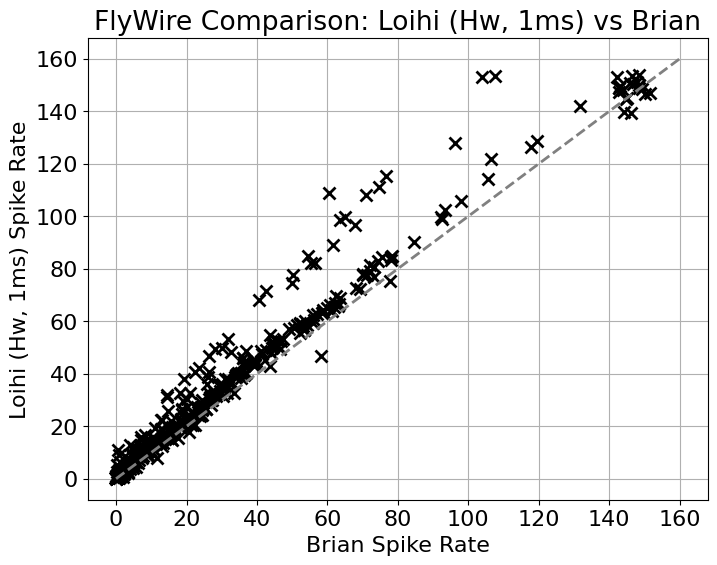}
    \caption{Comparing spike rates between Loihi 2 neuromorphic hardware implementation at 1ms timesteps to 0.1ms timesteps (left) and Brian 2 reference simulation (right)}
    \label{fig:comp_1mstimesteps}
\end{figure}

In addition to these model approximations, we also implemented a version of the FlyWire network model on Loihi 2 that increased the timestep size to 1ms. For this, we rounded both the synaptic delays and refractory periods to 2ms (2 timesteps) and adjusted the fixed-point constants of the neurons models accordingly. We collected spike rates over ten independent trials, and show the parity plots in Figure \ref{fig:comp_1mstimesteps} comparing this version of the model to the original timestep size of 0.1ms and to the Brian 2 reference simulation. We find that although the general effect of capped weights still persists, due to the increased integration interval, the effect from using conductance-only inputs was mitigated.

\subsection{Runtime Comparisons}

\begin{figure}[h!]
    \centering
    \includegraphics[width=0.7\linewidth]{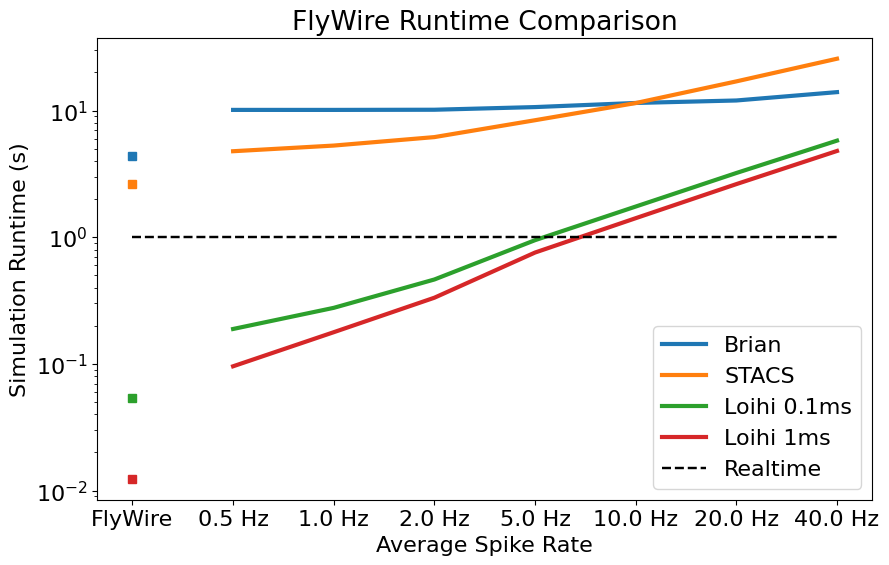} 
    \caption{Runtime performance comparison between different FlyWire implementations for the sugar neuron experiment and average baseline background spiking rates}
    \label{fig:performance_fig}
\end{figure}

\begin{figure}[h!]
    \centering
    \includegraphics[width=0.7\linewidth]{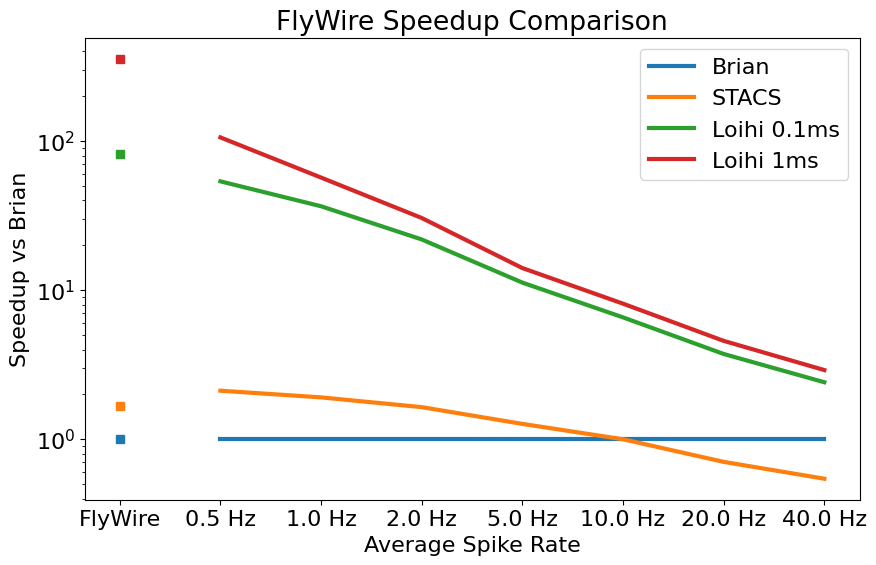} 
    \caption{Speedup compared to Brian 2 reference simulation between different FlyWire implementations for the sugar neuron experiment and average baseline background spiking rates}
    \label{fig:speedup_fig}
\end{figure}

In addition to validating the accuracy of our neuromorphic implementation, we also conducted some performance measurements for the sugar neuron experiment and our simple scaling study across different baseline spiking rates. The execution or wall clock time per 1s of simulated time was measured across the different simulators as well as for the Loihi 2 implementation at 0.1ms and 1ms timesteps (without spike probes). We collect our runtime results the accompanying speedup comparisons with respect to the Brian 2 reference simulation in Figures \ref{fig:performance_fig} and \ref{fig:speedup_fig}, respectively. We also compile the runtime results in Table \ref{tab:performance_tab}.

\begin{table}[h!]
    \centering
    \begin{tabular}{l|r|r|r|r|r|r|r|r}
        Platform & FlyWire (ms) & 0.5Hz (s) & 1.0Hz & 2.0Hz & 5.0Hz & 10.0Hz & 20.0Hz & 40.0Hz\\\hline
        Brian 2 & 4419 $\pm$ 236 & 10.13 & 10.13 & 10.16 & 10.66 & 11.49 & 12.02 & 13.99 \\\hline
        STACS & 2656 $\pm$ 79.8 & 4.778 & 5.296 & 6.177 & 8.396 & 11.47 & 17.01 & 25.68 \\\hline
        Loihi 2 (0.1ms) & 53.76 $\pm$ 0.896 & 0.189 & 0.277 & 0.464 & 0.948 & 1.748 & 3.216 & 5.793 \\\hline
        Loihi 2 (1ms) & 12.40 $\pm$ 0.279 & 0.0958 & 0.178 & 0.333 & 0.757 & 1.414 & 2.627 & 4.800
    \end{tabular}
    \caption{Runtime performance for 1s of simulated time between different FlyWire implementations for the sugar neuron experiment and average baseline background spiking rates}
    \label{tab:performance_tab}
\end{table}

Ideally neuromorphic hardware should be extremely well-suited for spiking neural network simulations, as the architecture is inspired by biological neural architectures. We found that for configurations with lower overall spiking activity, that the implementation on Loihi 2 was capable of performing better than realtime. Compared to the Brian 2 reference simulation, which has largely constant runtime performance regardless of spiking activity, we achieved speedups over 100x at sparser activity levels.

These results demonstrate that neuromorphic computing platforms such as Loihi 2 may be a powerful option for accelerating biological network simulations, though there are some challenges that must be overcome to take fuller advantage of the hardware. Furthermore, the unique characteristics of neuromorphic hardware presents a non-conventional trade-off space when reasoning about computation, communication, and especially memory. We expect some of the methods developed here for implementing the FlyWire network model onto Loihi 2 will also be useful for mapping other connectomes or neuromorphic algorithms in general.

\subsection{Network Compilation Process}

Although not directly related to network validation and performance, we also wanted to discuss some additional challenges and lessons learned from the FlyWire neuromorphic implementation effort. Primarily, the use of an intermediate representation and data structures that could translate between the global unified representation common to simulation and the local distributed representation necessary for hardware was indispensable to the conversion process, and will likely become even more important as we consider scaling up the size of our networks. Ideally, these processes for converting from simulation to hardware could be better integrated into a more automated and performant neuromorphic compilation toolchain.

One of the more important considerations in partitioning the network was keeping the resulting partitions within hardware limitations. On Loihi 2, this ended up being not just the total synaptic memory available, but also limits for the size of the axon programs and reserving enough space for the spike buffer. Although we used the amount of effective fan-in and fan-out per partition as a proxy to estimate the memory usage per neurocore, better heuristics to approximate or even calculate the expected usage prior to compilation through the NxCore API would have expedited the iterative partitioning process. Incorporating additional hardware information about the inter-core and inter-chip timing distances and available bandwidth could lead to even more sophisticated partitioning and mapping approaches.

Another aspect of the compilation process that could have been improved was parallelizing the construction and compilation of the network objects. Especially since the resulting partitions and the data structures are essentially self-contained with respect to their assigned neurocores, populating and writing out the data onto hardware should in theory be possible to execute as a multi-threaded process. Due to its multi-chip footprint, we observed that the compilation process of the FlyWire network was somewhat impeded by a number of unnecessary serial bottlenecks in the software infrastructure. While this was still manageable for the FlyWire network, we expect that these issues will become more apparent with scale, whether for larger network models or when duplicating smaller ones (e.g. for conducting parameter sweeps). Of course, we recognize that the Loihi 2 is predominantly meant to be a research chip, and that improvements to software infrastructure takes considerable effort to implement.

Another potential area for improvement, perhaps in a future iteration of the hardware, could be the support for more sophisticated communication schemes. In this work we explored strategies for compressing either the sending or receiving side connectivity information. We saw the most significant impact from compressing the synaptic delivery information, but at the expense of increased spike volume and mainly being suitable for network `inference' (static weights). The introduction of synapse programs to control the synaptic delivery (analogous to the current axon programs) could be useful for balancing between spike volume, memory requirements, and functionality, but perhaps just the option for one more layer of indirection to locally distribute unique axon indexes across a unique synapse list could be sufficient. Alternatively, if we expect that neuromorphic hardware continues or increases its bias toward efficient spiking communication (perhaps with optical techniques), then it may be more viable to introduce more complexity in the axon programs to incorporate sending side delays. That is, instead of relying on a shift buffer of dendritic accumulators on the receiving side to manage unique delays, we could buffer the outgoing spike on the sending side along with a staggered organization of axon programs to send spike messages to their targets only on the timestep that they would be processed.

\section{Conclusion}

Most notably, these results demonstrate that with only modest optimizations today's neuromorphic platforms can outperform conventional simulations in terms of time. As shown in Table \ref{tab:performance_tab}, the Loihi 2 simulation ran between {\fontfamily{ptm}\selectfont\textasciitilde}3x\textendash{\fontfamily{ptm}\selectfont\textasciitilde}350x faster than the reference Brian 2 simulation which is optimized for conventional simulations. While this is a single biological connectome, this observation supports projections that suggest that moving to mammalian-scale simulations of biologically realistic connectomes will likely require hardware such as neuromorphic~\cite{wang2024path}.  

While neuroscience has not fully embraced large-scale simulations in research and clinical settings, other biomedical fields increasingly have looked towards simulation to guide experimental and therapeutic strategies. While scientific discovery can perhaps get by with far greater than real-time simulation costs, for potential digital twin neurotechnology applications, such as targeting interventions for seizures, it will be increasingly essential that simulations can be performed more rapidly to tailor individualized surgical or pharmacological interventions.  This necessity of faster than real-time simulations is likely greater in neural simulations since the inherent complexity and stochasticity of neural systems will demand robust uncertainty quantification and sensitivity analyses, both of which will benefit from accelerating simulation times.

With these applications in mind, these results are suggestive that future neuromorphic hardware can impact neural simulations, but there are also opportunities to improve neural architectures for this purpose. In particular, the neuromorphic speed-up we observed is heavily dependent on the average activity of the model, with sparser activity leading to maximal speedups. This suggests that neuromorphic hardware may be well-suited for cortical simulations, but perhaps a hybrid conventional--neuromorphic strategy may be useful for simulations if certain high-activity regions (e.g., certain brain stem nuclei) are being modeled. Similarly, the connectivity diagram, in particular the fan-in of neurons in the connectome, is a critical consideration in mapping the circuit to neuromorphic hardware, and future designers may consider the full spectrum of biological connectivity. Other strategies may also make sense as more knowledge emerges from neuroscience field.

In conclusion, the results shown in this paper strongly support the long-standing hypothesized vision that neuromorphic hardware can be a powerful tool in full-scale neural simulations. While the Drosophila brain is a relatively simple neural system, it provides a circuit-level complexity far beyond most of today's ANNs and SNNs, demonstrating that the event-driven parallel design of today's spiking neuromorphic systems are well-suited for such circuits. The alignment of emerging neuromorphic hardware with emerging connectome resources represents a compelling opportunity for research going forward.

\section*{Acknowledgments}
We thank the Advanced Simulation and Computing program at the U.S. Department of Energy for supporting this research.

\insertfundingstatement{long}

\bibliographystyle{IEEEtran}
\bibliography{references}

\begin{thebibliography}{10}
\providecommand{\url}[1]{#1}
\csname url@samestyle\endcsname
\providecommand{\newblock}{\relax}
\providecommand{\bibinfo}[2]{#2}
\providecommand{\BIBentrySTDinterwordspacing}{\spaceskip=0pt\relax}
\providecommand{\BIBentryALTinterwordstretchfactor}{4}
\providecommand{\BIBentryALTinterwordspacing}{\spaceskip=\fontdimen2\font plus
\BIBentryALTinterwordstretchfactor\fontdimen3\font minus
  \fontdimen4\font\relax}
\providecommand{\BIBforeignlanguage}[2]{{%
\expandafter\ifx\csname l@#1\endcsname\relax
\typeout{** WARNING: IEEEtran.bst: No hyphenation pattern has been}%
\typeout{** loaded for the language `#1'. Using the pattern for}%
\typeout{** the default language instead.}%
\else
\language=\csname l@#1\endcsname
\fi
#2}}
\providecommand{\BIBdecl}{\relax}
\BIBdecl

\bibitem{dorkenwald2024neuronal}
S.~Dorkenwald, A.~Matsliah, A.~R. Sterling, P.~Schlegel, S.-C. Yu, C.~E.
  McKellar, A.~Lin, M.~Costa, K.~Eichler, Y.~Yin \emph{et~al.}, ``Neuronal
  wiring diagram of an adult brain,'' \emph{Nature}, vol. 634, no. 8032, pp.
  124--138, 2024.

\bibitem{lin2024network}
A.~Lin, R.~Yang, S.~Dorkenwald, A.~Matsliah, A.~R. Sterling, P.~Schlegel, S.-c.
  Yu, C.~E. McKellar, M.~Costa, K.~Eichler \emph{et~al.}, ``Network statistics
  of the whole-brain connectome of drosophila,'' \emph{Nature}, vol. 634, no.
  8032, pp. 153--165, 2024.

\bibitem{schlegel2024whole}
P.~Schlegel, Y.~Yin, A.~S. Bates, S.~Dorkenwald, K.~Eichler, P.~Brooks, D.~S.
  Han, M.~Gkantia, M.~Dos~Santos, E.~J. Munnelly \emph{et~al.}, ``Whole-brain
  annotation and multi-connectome cell typing of drosophila,'' \emph{Nature},
  vol. 634, no. 8032, pp. 139--152, 2024.

\bibitem{denk2012structural}
W.~Denk, K.~L. Briggman, and M.~Helmstaedter, ``Structural neurobiology:
  missing link to a mechanistic understanding of neural computation,''
  \emph{Nature Reviews Neuroscience}, vol.~13, no.~5, pp. 351--358, 2012.

\bibitem{sporns2005human}
O.~Sporns, G.~Tononi, and R.~K{\"o}tter, ``The human connectome: a structural
  description of the human brain,'' \emph{PLoS computational biology}, vol.~1,
  no.~4, p. e42, 2005.

\bibitem{bargmann2013connectome}
C.~I. Bargmann and E.~Marder, ``From the connectome to brain function,''
  \emph{Nature methods}, vol.~10, no.~6, pp. 483--490, 2013.

\bibitem{shiu2024drosophila}
P.~K. Shiu, G.~R. Sterne, N.~Spiller, R.~Franconville, A.~Sandoval, J.~Zhou,
  N.~Simha, C.~H. Kang, S.~Yu, J.~S. Kim \emph{et~al.}, ``A drosophila
  computational brain model reveals sensorimotor processing,'' \emph{Nature},
  vol. 634, no. 8032, pp. 210--219, 2024.

\bibitem{hbpnmc2013}
A.~Calimera, E.~Macii, and M.~Poncino, ``The human brain project and
  neuromorphic computing,'' \emph{Functional neurology}, vol.~28, pp. 191--196,
  10 2013.

\bibitem{lu2024simulation}
W.~Lu, X.~Du, J.~Wang, L.~Zeng, L.~Ye, S.~Xiang, Q.~Zheng, J.~Zhang, N.~Xu,
  J.~Feng \emph{et~al.}, ``Simulation and assimilation of the digital human
  brain,'' \emph{Nature Computational Science}, pp. 1--9, 2024.

\bibitem{wang2024path}
F.~Wang and J.~B. Aimone, ``On the path toward brain-scale simulations,''
  \emph{Nature Computational Science}, pp. 1--2, 2024.

\bibitem{kudithipudi2025neuromorphic}
D.~Kudithipudi, C.~Schuman, C.~M. Vineyard, T.~Pandit, C.~Merkel, R.~Kubendran,
  J.~B. Aimone, G.~Orchard, C.~Mayr, R.~Benosman \emph{et~al.}, ``Neuromorphic
  computing at scale,'' \emph{Nature}, vol. 637, no. 8047, pp. 801--812, 2025.

\bibitem{knight2016synapse}
J.~C. Knight and S.~B. Furber, ``Synapse-centric mapping of cortical models to
  the spinnaker neuromorphic architecture,'' \emph{Frontiers in neuroscience},
  vol.~10, p. 420, 2016.

\bibitem{sharp2012power}
T.~Sharp, F.~Galluppi, A.~Rast, and S.~Furber, ``Power-efficient simulation of
  detailed cortical microcircuits on spinnaker,'' \emph{Journal of neuroscience
  methods}, vol. 210, no.~1, pp. 110--118, 2012.

\bibitem{imam2020rapid}
N.~Imam and T.~A. Cleland, ``Rapid online learning and robust recall in a
  neuromorphic olfactory circuit,'' \emph{Nature Machine Intelligence}, vol.~2,
  no.~3, pp. 181--191, 2020.

\bibitem{hassabis2017neuroai}
D.~Hassabis, D.~Kumaran, C.~Summerfield, and M.~Botvinick,
  ``Neuroscience-inspired artificial intelligence,'' \emph{Neuron}, vol.~95,
  no.~2, pp. 245--258, 2017.

\bibitem{resnet2016}
K.~He, X.~Zhang, S.~Ren, and J.~Sun, ``Deep residual learning for image
  recognition,'' in \emph{2016 IEEE Conference on Computer Vision and Pattern
  Recognition (CVPR)}, 2016, pp. 770--778.

\bibitem{attention2017}
A.~Vaswani, N.~Shazeer, N.~Parmar, J.~Uszkoreit, L.~Jones, A.~N. Gomez,
  L.~Kaiser, and I.~Polosukhin, ``Attention is all you need,'' in
  \emph{Proceedings of the 31st International Conference on Neural Information
  Processing Systems}, ser. NIPS'17.\hskip 1em plus 0.5em minus 0.4em\relax Red
  Hook, NY, USA: Curran Associates Inc., 2017, p. 6000–6010.

\bibitem{goodfellow2016dl}
I.~Goodfellow, Y.~Bengio, and A.~Courville, \emph{Deep Learning}.\hskip 1em
  plus 0.5em minus 0.4em\relax MIT Press, 2016,
  \url{http://www.deeplearningbook.org}.

\bibitem{dlsurvey2021}
\BIBentryALTinterwordspacing
L.~Alzubaidi, J.~Zhang, A.~J. Humaidi, A.~Al-Dujaili, Y.~Duan, O.~Al-Shamma,
  J.~Santamar{\'\i}a, M.~A. Fadhel, M.~Al-Amidie, and L.~Farhan, ``Review of
  deep learning: concepts, cnn architectures, challenges, applications, future
  directions,'' \emph{Journal of Big Data}, vol.~8, no.~1, p.~53, 2021.
  [Online]. Available: \url{https://doi.org/10.1186/s40537-021-00444-8}
\BIBentrySTDinterwordspacing

\bibitem{nas2019survey}
T.~Elsken, J.~H. Metzen, and F.~Hutter, ``Neural architecture search: a
  survey,'' \emph{J. Mach. Learn. Res.}, vol.~20, no.~1, p. 1997–2017, Jan.
  2019.

\bibitem{schuman2020eons}
\BIBentryALTinterwordspacing
C.~D. Schuman, J.~P. Mitchell, R.~M. Patton, T.~E. Potok, and J.~S. Plank,
  ``Evolutionary optimization for neuromorphic systems,'' in \emph{Proceedings
  of the 2020 Annual Neuro-Inspired Computational Elements Workshop}, ser. NICE
  '20.\hskip 1em plus 0.5em minus 0.4em\relax New York, NY, USA: Association
  for Computing Machinery, 2020. [Online]. Available:
  \url{https://doi.org/10.1145/3381755.3381758}
\BIBentrySTDinterwordspacing

\bibitem{schuman2024hairball}
\BIBentryALTinterwordspacing
C.~D. Schuman, C.~P. Rizzo, G.~S. Rose, and J.~S. Plank, ``Embracing the
  hairball: An investigation of recurrence in spiking neural networks for
  control,'' in \emph{NICE: Neuro-Inspired Computational Elements Workshop},
  April 2024. [Online]. Available:
  \url{https://ieeexplore.ieee.org/document/10548512/}
\BIBentrySTDinterwordspacing

\bibitem{white1986structure}
J.~G. White, E.~Southgate, J.~N. Thomson, S.~Brenner \emph{et~al.}, ``The
  structure of the nervous system of the nematode caenorhabditis elegans,''
  \emph{Philos Trans R Soc Lond B Biol Sci}, vol. 314, no. 1165, pp. 1--340,
  1986.

\bibitem{varshney2011celegans}
\BIBentryALTinterwordspacing
L.~R. Varshney, B.~L. Chen, E.~Paniagua, D.~H. Hall, and D.~B. Chklovskii,
  ``Structural properties of the caenorhabditis elegans neuronal network,''
  \emph{PLOS Computational Biology}, vol.~7, no.~2, pp. 1--21, 02 2011.
  [Online]. Available: \url{https://doi.org/10.1371/journal.pcbi.1001066}
\BIBentrySTDinterwordspacing

\bibitem{ahrens2013zebrafish}
\BIBentryALTinterwordspacing
M.~B. Ahrens, M.~B. Orger, D.~N. Robson, J.~M. Li, and P.~J. Keller,
  ``Whole-brain functional imaging at cellular resolution using light-sheet
  microscopy,'' \emph{Nature Methods}, vol.~10, no.~5, pp. 413--420, 2013.
  [Online]. Available: \url{https://doi.org/10.1038/nmeth.2434}
\BIBentrySTDinterwordspacing

\bibitem{microns2025functional}
J.~A. Bae \emph{et~al.}, ``Functional connectomics spanning multiple areas of
  mouse visual cortex,'' \emph{Nature}, vol. 640, no. 8058, pp. 435--447, 2025.

\bibitem{AllenInstitute_v1}
\BIBentryALTinterwordspacing
A.~Institute, ``Mv1 all layers,'' 2025, accessed: 2025-02-03. [Online].
  Available: \url{https://portal.brain-map.org/explore/models/mv1-all-layers}
\BIBentrySTDinterwordspacing

\bibitem{billeh2020systematic}
Y.~N. Billeh, B.~Cai, S.~L. Gratiy, K.~Dai, R.~Iyer, N.~W. Gouwens,
  R.~Abbasi-Asl, X.~Jia, J.~H. Siegle, S.~R. Olsen \emph{et~al.}, ``Systematic
  integration of structural and functional data into multi-scale models of
  mouse primary visual cortex,'' \emph{Neuron}, vol. 106, no.~3, pp. 388--403,
  2020.

\bibitem{potjans2012micro}
\BIBentryALTinterwordspacing
T.~C. Potjans and M.~Diesmann, ``The cell-type specific cortical microcircuit:
  Relating structure and activity in a full-scale spiking network model,''
  \emph{Cerebral Cortex}, vol.~24, no.~3, pp. 785--806, 12 2012. [Online].
  Available: \url{https://doi.org/10.1093/cercor/bhs358}
\BIBentrySTDinterwordspacing

\bibitem{awile2022modernizing}
O.~Awile, P.~Kumbhar, N.~Cornu, S.~Dura-Bernal, J.~G. King, O.~Lupton,
  I.~Magkanaris, R.~A. McDougal, A.~J. Newton, F.~Pereira \emph{et~al.},
  ``Modernizing the neuron simulator for sustainability, portability, and
  performance,'' \emph{Frontiers in Neuroinformatics}, vol.~16, p. 884046,
  2022.

\bibitem{goodman2009brian}
D.~F. Goodman and R.~Brette, ``The brian simulator,'' \emph{Frontiers in
  neuroscience}, vol.~3, p. 643, 2009.

\bibitem{stimberg2019brian}
M.~Stimberg, R.~Brette, and D.~F. Goodman, ``Brian 2, an intuitive and
  efficient neural simulator,'' \emph{elife}, vol.~8, p. e47314, 2019.

\bibitem{knight2018gpus}
J.~C. Knight and T.~Nowotny, ``Gpus outperform current hpc and neuromorphic
  solutions in terms of speed and energy when simulating a highly-connected
  cortical model,'' \emph{Frontiers in neuroscience}, vol.~12, p. 427264, 2018.

\bibitem{knight2021larger}
------, ``Larger gpu-accelerated brain simulations with procedural
  connectivity,'' \emph{Nature Computational Science}, vol.~1, no.~2, pp.
  136--142, 2021.

\bibitem{diesmann2001nest}
M.~Diesmann and M.-O. Gewaltig, ``Nest: An environment for neural systems
  simulations,'' \emph{Forschung und wisschenschaftliches Rechnen, Beitr{\"a}ge
  zum Heinz-Billing-Preis}, vol.~58, pp. 43--70, 2001.

\bibitem{wang2024scaling}
F.~Wang, S.~Kulkarni, B.~Theilman, F.~Rothganger, C.~Schuman, S.-H. Lim, and
  J.~B. Aimone, ``Scaling neural simulations in {STACS},'' \emph{Neuromorphic
  Computing and Engineering}, vol.~4, no.~2, p. 024002, 2024.

\bibitem{furber2020spinnaker}
S.~Furber and P.~Bogdan, \emph{Spinnaker-a spiking neural network
  architecture}.\hskip 1em plus 0.5em minus 0.4em\relax Now publishers, 2020.

\bibitem{furber2014spinnaker}
S.~B. Furber, F.~Galluppi, S.~Temple, and L.~A. Plana, ``The spinnaker
  project,'' \emph{Proceedings of the IEEE}, vol. 102, no.~5, pp. 652--665,
  2014.

\bibitem{schemmel2010wafer}
J.~Schemmel, D.~Br{\"u}derle, A.~Gr{\"u}bl, M.~Hock, K.~Meier, and S.~Millner,
  ``A wafer-scale neuromorphic hardware system for large-scale neural
  modeling,'' in \emph{2010 ieee international symposium on circuits and
  systems (iscas)}.\hskip 1em plus 0.5em minus 0.4em\relax IEEE, 2010, pp.
  1947--1950.

\bibitem{schmidt2023clean}
H.~Schmidt, J.~Montes, A.~Gr{\"u}bl, M.~G{\"u}ttler, D.~Husmann, J.~Ilmberger,
  J.~Kaiser, C.~Mauch, E.~M{\"u}ller, L.~Sterzenbach \emph{et~al.}, ``From
  clean room to machine room: commissioning of the first-generation brainscales
  wafer-scale neuromorphic system,'' \emph{Neuromorphic Computing and
  Engineering}, vol.~3, no.~3, p. 034013, 2023.

\bibitem{furber2016large}
S.~Furber, ``Large-scale neuromorphic computing systems,'' \emph{Journal of
  neural engineering}, vol.~13, no.~5, p. 051001, 2016.

\bibitem{gonzalez2024spinnaker2}
H.~A. Gonzalez, J.~Huang, F.~Kelber, K.~K. Nazeer, T.~Langer, C.~Liu,
  M.~Lohrmann, A.~Rostami, M.~Sch{\"o}ne, B.~Vogginger \emph{et~al.},
  ``Spinnaker2: A large-scale neuromorphic system for event-based and
  asynchronous machine learning,'' \emph{arXiv preprint arXiv:2401.04491},
  2024.

\bibitem{merolla2014million}
P.~A. Merolla, J.~V. Arthur, R.~Alvarez-Icaza, A.~S. Cassidy, J.~Sawada,
  F.~Akopyan, B.~L. Jackson, N.~Imam, C.~Guo, Y.~Nakamura \emph{et~al.}, ``A
  million spiking-neuron integrated circuit with a scalable communication
  network and interface,'' \emph{Science}, vol. 345, no. 6197, pp. 668--673,
  2014.

\bibitem{davies2018loihi}
M.~Davies, N.~Srinivasa, T.-H. Lin, G.~Chinya, Y.~Cao, S.~H. Choday, G.~Dimou,
  P.~Joshi, N.~Imam, S.~Jain \emph{et~al.}, ``Loihi: A neuromorphic manycore
  processor with on-chip learning,'' \emph{Ieee Micro}, vol.~38, no.~1, pp.
  82--99, 2018.

\bibitem{aimone2019composing}
J.~B. Aimone, W.~Severa, and C.~M. Vineyard, ``Composing neural algorithms with
  fugu,'' in \emph{Proceedings of the International Conference on Neuromorphic
  Systems}, 2019, pp. 1--8.

\bibitem{rhodes2019spinncolumn}
O.~Rhodes, L.~Peres, A.~G.~D. Rowley, A.~Gait, L.~A. Plana, C.~Brenninkmeijer,
  and S.~B. Furber, ``Real-time cortical simulation on neuromorphic hardware,''
  \emph{Philosophical Transactions of the Royal Society A: Mathematical,
  Physical and Engineering Sciences}, vol. 378, no. 2164, p. 20190160, 2020.

\bibitem{rathi2023survey}
\BIBentryALTinterwordspacing
N.~Rathi, I.~Chakraborty, A.~Kosta, A.~Sengupta, A.~Ankit, P.~Panda, and
  K.~Roy, ``Exploring neuromorphic computing based on spiking neural networks:
  Algorithms to hardware,'' \emph{ACM Comput. Surv.}, vol.~55, no.~12, Mar.
  2023. [Online]. Available: \url{https://doi.org/10.1145/3571155}
\BIBentrySTDinterwordspacing

\bibitem{schmidt2025brianscales}
H.~Schmidt, A.~Grübl, J.~Montes, E.~Müller, S.~Schmitt, and J.~Schemmel,
  ``Demonstrating the advantages of analog wafer-scale neuromorphic hardware,''
  in \emph{2025 Neuro Inspired Computational Elements (NICE)}, 2025, pp. 1--5.

\bibitem{severa2025benchmarking}
W.~Severa, F.~Wang, Y.~Ho, F.~Rothganger, A.~Daram, and E.~Gonzalez,
  ``Benchmarking spiking network partitioning methods on loihi 2,'' in
  \emph{Proceedings of the Great Lakes Symposium on VLSI 2025}, 2025, pp.
  898--904.

\bibitem{davies2021advancing}
M.~Davies, A.~Wild, G.~Orchard, Y.~Sandamirskaya, G.~A.~F. Guerra, P.~Joshi,
  P.~Plank, and S.~R. Risbud, ``Advancing neuromorphic computing with loihi: A
  survey of results and outlook,'' \emph{Proceedings of the IEEE}, vol. 109,
  no.~5, pp. 911--934, 2021.

\bibitem{shrestha2024efficient}
S.~B. Shrestha, J.~Timcheck, P.~Frady, L.~Campos-Macias, and M.~Davies,
  ``Efficient video and audio processing with loihi 2,'' in \emph{ICASSP
  2024-2024 IEEE International Conference on Acoustics, Speech and Signal
  Processing (ICASSP)}.\hskip 1em plus 0.5em minus 0.4em\relax IEEE, 2024, pp.
  13\,481--13\,485.

\bibitem{abreu2025neuromorphic}
S.~Abreu, S.~B. Shrestha, R.-J. Zhu, and J.~Eshraghian, ``Neuromorphic
  principles for efficient large language models on intel loihi 2,''
  \emph{arXiv preprint arXiv:2503.18002}, 2025.

\bibitem{vineyard2019low}
C.~M. Vineyard, R.~Dellana, J.~B. Aimone, F.~Rothganger, and W.~M. Severa,
  ``Low-power deep learning inference using the spinnaker neuromorphic
  platform,'' in \emph{Proceedings of the 7th Annual Neuro-inspired
  computational elements Workshop}, 2019, pp. 1--7.

\bibitem{kelber2020mapping}
F.~Kelber, B.~Wu, B.~Vogginger, J.~Partzsch, C.~Liu, M.~Stolba, and C.~Mayr,
  ``Mapping deep neural networks on spinnaker2,'' in \emph{Proceedings of the
  2020 Annual Neuro-Inspired Computational Elements Workshop}, 2020, pp. 1--3.

\bibitem{rostami2022prop}
A.~Rostami, B.~Vogginger, Y.~Yan, and C.~G. Mayr, ``E-prop on spinnaker 2:
  Exploring online learning in spiking rnns on neuromorphic hardware,''
  \emph{Frontiers in Neuroscience}, vol.~16, p. 1018006, 2022.

\bibitem{nazeer2024language}
K.~K. Nazeer, M.~Sch{\"o}ne, R.~Mukherji, B.~Vogginger, C.~Mayr, D.~Kappel, and
  A.~Subramoney, ``Language modeling on a spinnaker2 neuromorphic chip,'' in
  \emph{2024 IEEE 6th International Conference on AI Circuits and Systems
  (AICAS)}.\hskip 1em plus 0.5em minus 0.4em\relax IEEE, 2024, pp. 492--496.

\bibitem{alom2018deep}
M.~Z. Alom, T.~Josue, M.~N. Rahman, W.~Mitchell, C.~Yakopcic, and T.~M. Taha,
  ``Deep versus wide convolutional neural networks for object recognition on
  neuromorphic system,'' in \emph{2018 International Joint Conference on Neural
  Networks (IJCNN)}.\hskip 1em plus 0.5em minus 0.4em\relax IEEE, 2018, pp.
  1--8.

\bibitem{schmitt2017neuromorphic}
S.~Schmitt, J.~Kl{\"a}hn, G.~Bellec, A.~Gr{\"u}bl, M.~Guettler, A.~Hartel,
  S.~Hartmann, D.~Husmann, K.~Husmann, S.~Jeltsch \emph{et~al.}, ``Neuromorphic
  hardware in the loop: Training a deep spiking network on the brainscales
  wafer-scale system,'' in \emph{2017 international joint conference on neural
  networks (IJCNN)}.\hskip 1em plus 0.5em minus 0.4em\relax IEEE, 2017, pp.
  2227--2234.

\bibitem{akopyan2015truenorth}
F.~Akopyan, J.~Sawada, A.~Cassidy, R.~Alvarez-Icaza, J.~Arthur, P.~Merolla,
  N.~Imam, Y.~Nakamura, P.~Datta, G.-J. Nam \emph{et~al.}, ``Truenorth: Design
  and tool flow of a 65 mw 1 million neuron programmable neurosynaptic chip,''
  \emph{IEEE transactions on computer-aided design of integrated circuits and
  systems}, vol.~34, no.~10, pp. 1537--1557, 2015.

\bibitem{loihi2techbrief}
Intel, ``Taking neuromorphic computing to the next level with loihi 2,'' Intel,
  Tech. Rep., September 2021.

\bibitem{wang2015simulation}
F.~Wang, ``Simulation tool for asynchronous cortical streams ({STACS}):
  interfacing with spiking neural networks,'' \emph{Procedia Computer Science},
  vol.~61, pp. 322--327, 2015.

\end{thebibliography}

\appendix


\end{document}